\documentclass{JHEP3}

\usepackage{xspace}
\usepackage{amsmath}
\usepackage{epsfig}

\newcommand{\Tr}{\mathop{\rm Tr}\nolimits}
\newcommand{\SU}{\mathop{\rm SU}\nolimits}
\newcommand{\U}{\mathop{\rm {}U}\nolimits}
\newcommand{\SO}{\mathop{\rm SO}\nolimits}
\def\OO{{\mathrm O}}
\newcommand{\USp}{\mathop{\rm {}USp}\nolimits}

\newcommand{\hw}{Ho\v{r}ava-Witten\xspace}
\newcommand{\half}{\ensuremath{\frac{1}{2}}}
\newcommand{\qtr}{\frac{1}{4}}
\newcommand{\tg}{{\tt{g}}}
\newcommand{\LL}{\mathcal{L}}
\newcommand{\HH}{\ensuremath{\mathcal{H}}}
\newcommand{\M}{\ensuremath{\mathcal{M}}}
\newcommand{\ch}{${\cal {H}}\ $}
\newcommand{\beq}{\begin{equation}}
\newcommand{\eeq}{\end{equation}}
\newcommand{\beqa}{\begin{eqnarray}}
\newcommand{\eeqa}{\end{eqnarray}}
\newcommand{\ba}{\begin{array}}
\newcommand{\ea}{\end{array}}
\newcommand{\vac}{|0\rangle}
\newcommand{\OOO}{{\mathcal O}}

\def\IZ{{\mathbb{Z}}}
\def\IR{{\mathbb{R}}}

\def\rep{representation}
\def\su41{$\SU(4|1)$}
\def\hw{highest weight }

\preprint{{\tt hep-th/0306051 v2}\\
HUTP-03/A040\\
SU-ITP-03/10\\
HEP-UK-0018
}

\title{Heterotic plane wave matrix models\\ and giant gluons}

\author{Lubo\v{s} Motl\\
Jefferson Physical Laboratory, Harvard University\\
Cambridge, MA 02138, USA\\
E-mail: \email{motl@feynman.harvard.edu}}

\author{Andrew Neitzke\\
Jefferson Physical Laboratory, Harvard University\\
Cambridge, MA 02138, USA\\
E-mail: \email{neitzke@fas.harvard.edu}}

\author{Mohammad M. Sheikh-Jabbari\\
Department of Physics, Stanford University\\
382 via Pueblo Mall, Stanford, CA 94305-4060, USA\\
E-mail: \email{jabbari@itp.stanford.edu}}

\abstract{In this paper we define and study a matrix model describing
the M-theory plane wave background with a single Ho\v{r}ava-Witten domain wall. In
the limit of infinite $\mu$, the matrix model action becomes
quadratic and we can identify the matrix Hamiltonian with a regularized
Hamiltonian for hemispherical membranes that carry fermionic degrees of
freedom on their boundaries. The number of fermionic degrees
of freedom must be sixteen; this
condition arises naturally in the framework of the
matrix model.  We can also prove the exact $E_8$ symmetry of
the spectrum around the membrane vacua at infinite $\mu$, which arises as a current algebra
at level one just as in the heterotic string. 
We also find the full $E_8$ gauge multiplet as well as the multiple-gluon states, carried
by collections of hemispherical membranes.  Finally we
discuss the dual description of the hemispherical membranes in terms of
spherical fivebranes immersed in the domain wall; we identify the correct vacuum of
the matrix model and make some preliminary remarks about comparison with
the $(1,0)$ superconformal field theory.}


\begin{document}

\section{Introduction}\label{intro}

Although 11-dimensional M-theory seemed to be hidden in a cloud of magic
and mystery right after it was discovered \cite{wittenvarious}, it has
become the first background of supersymmetric quantum gravity 
for which we can describe the dynamics in a
fully nonperturbative framework. Banks, Fischler, Shenker and Susskind
\cite{bfss} realized in 1996 that the reduction of
9+1-dimensional $\U(N)$ Super Yang-Mills theory to 0+1 dimensions 
not only describes the low-energy dynamics
of nonrelativistic D0-branes and of a single discretized
supermembrane \cite{nicolai}, but is also capable of giving a
quantitative answer to an
arbitrary dynamical question about the sector of M-theory quantized in
DLCQ (Discrete Light Cone Quantization) with $N=p^+ R$ units of the
light-like longitudinal momentum \cite{SussDLCQ}.

In the matrix theory picture, 
the gravity multiplet carrying $k$ units of the longitudinal momentum is
described as a bound state of $k$ D0-branes.  Arguments from string duality show
that a unique such
bound state should exist for every positive integer $k$.  The most complete
direct argument in favor of its existence has been given in the case of
the $\SU(2)$ matrix model \cite{sethist} and some evidence has also been
given for prime integer values of $k$ \cite{prime}. It is however
generally believed that such a bound state must exist for each $k$, for the
following reason.
The matrix model describing 11-dimensional M-theory
can be continuously connected to other matrix models which describe its
compactifications. In particular there is a 1+1-dimensional matrix model 
that describes
compactification to 10 dimensions which can be solved at weak coupling.
This matrix model can thus be shown to describe
perturbative type IIA string theory \cite{lumodvv,bsdvv,dvv}
within a consistent non-perturbative framework.  Perturbative
type IIA strings are easily shown to carry a supergravity multiplet.
This state can be continued into strong coupling, and therefore we can
argue that the BFSS model itself contains the required states representing
the graviton supermultiplet.

\EPSFIGURE{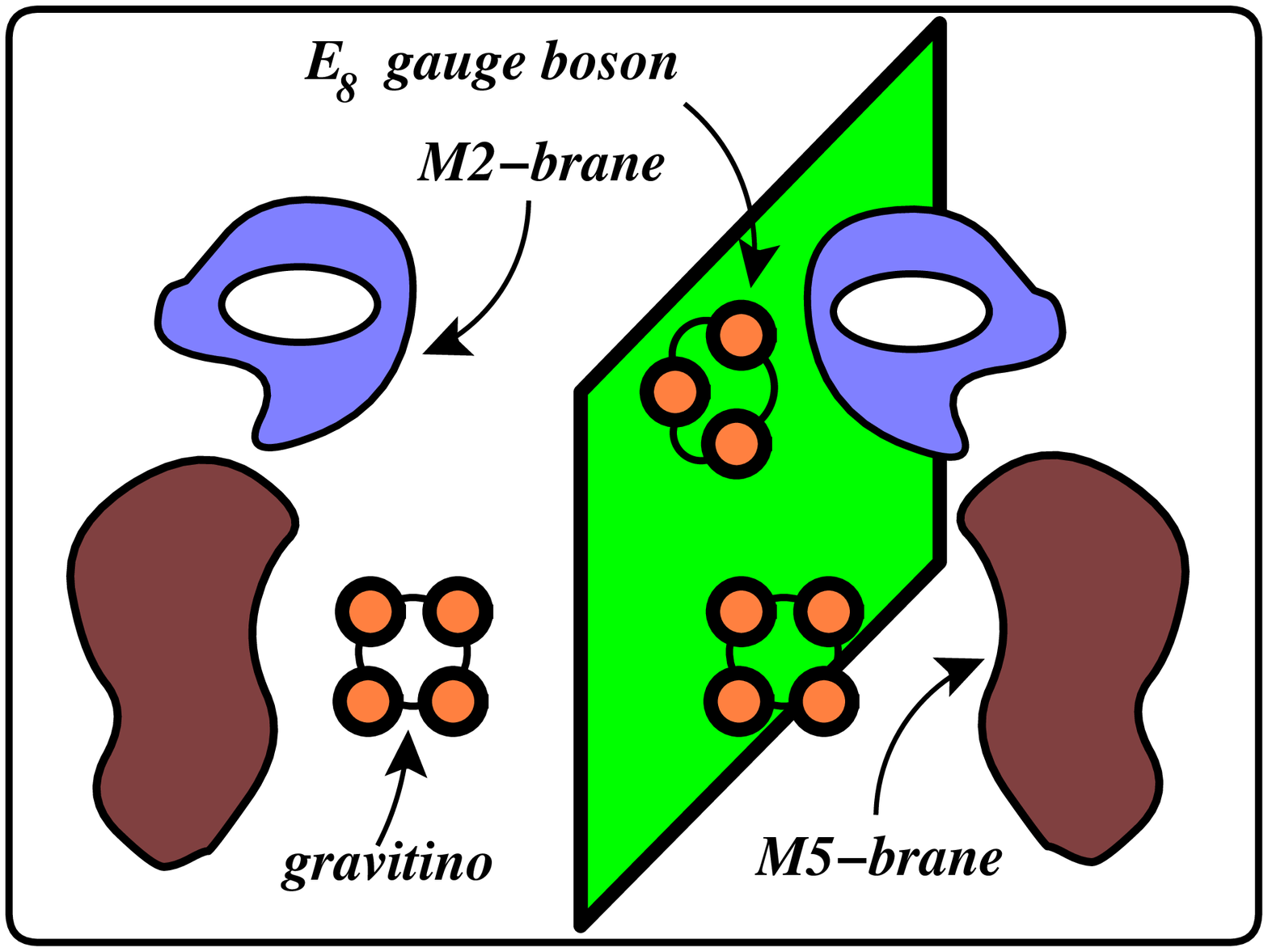,width=85mm}{A typical state
in heterotic M-theory. One gravitino and one gauge boson
are depicted as bound states of D0-branes.
\label{scifi}}

$\!$More complicated dynamics arise when we consider the formal $\IZ_2$
orientifold of the BFSS model, namely the $\OO(N)$ heterotic matrix model
\cite{danielsson,motlquat,kimrey,bss,banksmotl,lowehet,reyhet,horavamatrix}.
In the $\IZ_2$ quotient of M-theory in 11 dimensions, 
we find not only the gravity multiplet but also 
a gauge multiplet confined to the 
fixed locus of the $\IZ_2$ 
orientifold, otherwise known as the Ho\v{r}ava-Witten
domain wall \cite{howittenone,howittentwo}.  So there should be appropriate
\linebreak
bound
states in the quantum mechanics describing these states as well.

A separate line of reasoning \cite{bmn} led to the discovery of a massive
deformation of the BFSS matrix model.  The deformed model describes M-theory on a
plane wave background which arises as the Penrose
limit of $AdS_4 \times S^7$ or of
$AdS_7\times S^4$ (they turn out to be identical).  One major advantage
of this model is that the usual difficulties in matrix theory associated with finding the bound 
state spectrum are absent; the mass deformation lifts all the flat directions and so 
the states corresponding to the graviton supermultiplet as well as the
multi-graviton states are easy to find.  In fact, they are
localized near the ``fuzzy sphere'' classical configurations which
are related to well-known giant gravitons \cite{giantg}.  In the limit of
infinite background flux $\mu$, the action becomes quadratic and the 
theory can be solved perturbatively in $1/ \mu$.

\EPSFIGURE{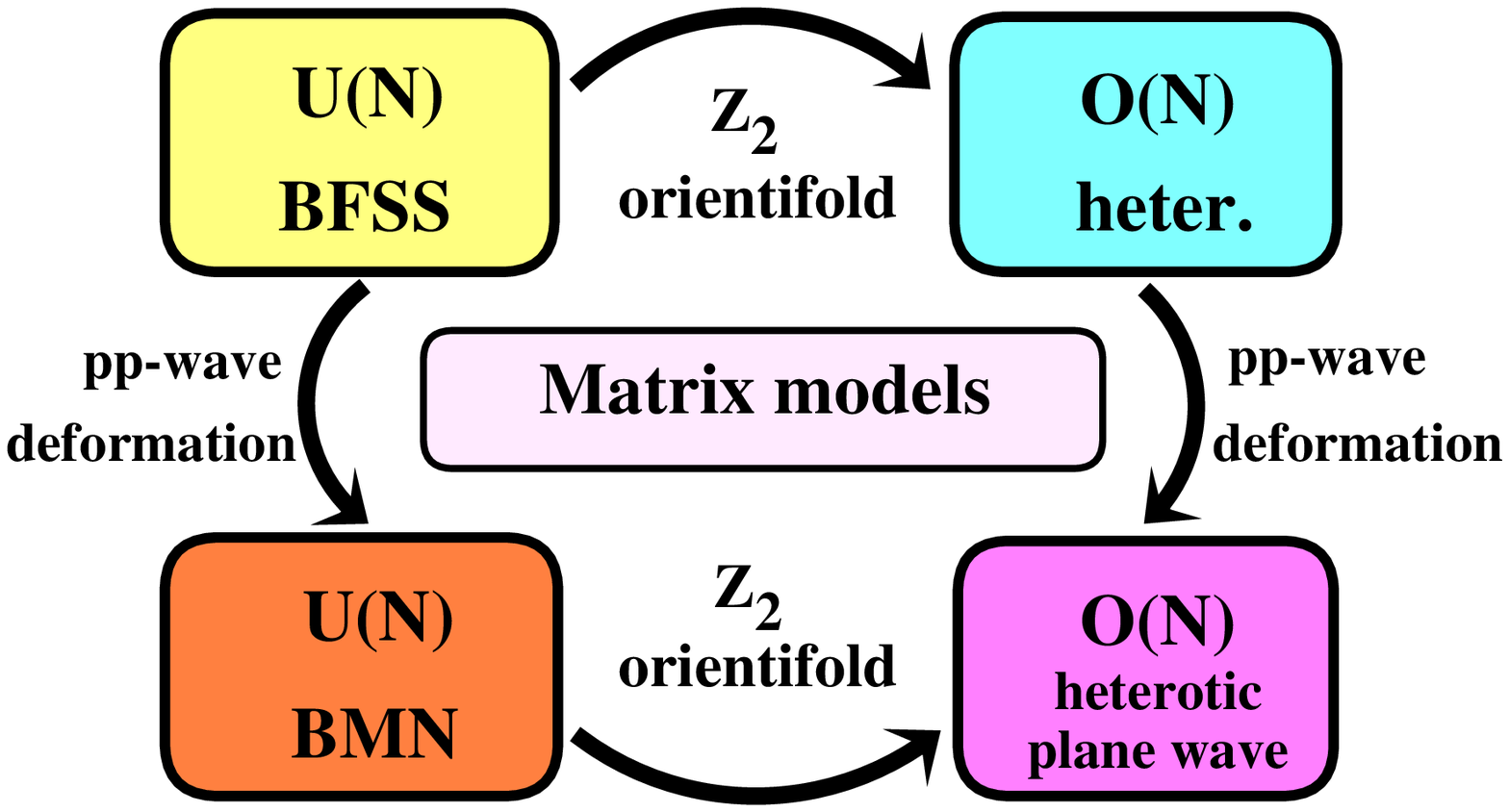,width=85mm}{A commutative diagram showing
the flat space BFSS Matrix theory, its $\IZ_2$ orientifold, and their
pp-wave massive deformations.
\label{commutdiag}}

In this paper we combine the 
\linebreak
methods of heterotic matrix models with those
of pp-wave matrix models by taking a $\IZ_2$ quotient of the BMN matrix
model.  We will find that the massive deformation brings the same advantage
here that it brought in the maximally supersymmetric case:  
\linebreak
namely, we
will be able to see the vector supermultiplet carried by classical ``giant
gluons,'' which are hemispherical membranes with fermionic degrees of
freedom supported on the boundary.

The paper \cite{tfivebrane} also presented evidence that in
the BMN matrix model a vacuum corresponding to a large 
number of very small spherical membranes 
admits a dual interpretation as 
a spherical transverse fivebrane---an object which had previously
eluded detection in the matrix theory framework \cite{branesfrommatrices}.  In our
model we will find a
corresponding fivebrane vacuum; it is a candidate to describe a spherical fivebrane
immersed in the domain wall.  We will give some evidence for this identification.

\section{Brief review of the BMN matrix model}

We begin with the maximally supersymmetric
plane wave background of 11-dimensional supergravity,
which can be obtained as a Penrose limit of $AdS_4 \times 
S^7$ or of $AdS_7 \times S^4$ \cite{BlauPen}:
\begin{align}\label{planewave}
ds^2 &= -2 dx^+ dx^- + \sum_{A=1}^9 dx^A dx^A - \left(\sum_{i=1}^3 \left(\frac{\mu}{3}\right)^2 x^i x^i  + \sum_{a=4}^9 \left(\frac{\mu}{6}\right)^2 x^a x^a \right) dx^+ dx^+, \\
F_{123+} &= \mu.
\end{align}
Berenstein, Maldacena, and Nastase \cite{bmn} proposed a supersymmetric
matrix model describing discrete light-cone quantization (DLCQ) of M-theory in the
background \eqref{planewave}.
Its Hamiltonian can be thought of as a massive
deformation of the BFSS matrix model \cite{bfss}
\begin{equation}
H = H_0 + H_\mu \equiv p^-.
\end{equation}
Here $H_0$ is the BFSS Hamiltonian 
describing DLCQ of M-theory in 11 flat dimensions,
\begin{equation}
H_0 = R \Tr \left( \half \Pi_A^2 - \qtr [X^A, X^B]^2 - \half \Psi^t 
\gamma^A [X^A, \Psi] \right)
\end{equation}
and $H_\mu$ is the massive deformation
\begin{equation}
H_\mu = \frac{R}{2} \Tr \left( \left(\frac{\mu}{3R}\right)^2 
(X^i)^2 + \left(\frac{\mu}{6 R}\right)^2 (X^a)^2 + 
i \frac{\mu}{4 R} \Psi^T \gamma^{123} \Psi + i \frac{2 \mu}{3 R} 
\epsilon^{ijk} X^i X^j X^k \right).
\end{equation}
Here $\Psi$ is a set of $16$ Grassmann-valued Hermitian
matrices transforming as the $16$-component real 
spinor of $\SO(9)$, 
$i$ runs from $1$ to $3$, $a$ runs from 4 to 9, and $A$ runs from $1$ to 
$9$.

This Hamiltonian has symmetry supergroup $\SU(4|2)$ and it is
convenient to write it in a way which makes that manifest. For the bosons
we have done this already by splitting the indices into $i$ and $a$.  For
the fermions a convenient formalism appeared in \cite{matrixperturb}:  
under
\begin{align}
\SO(9) & \to \SO(6) \times \SO(3)
\intertext{the fermions split as}
\bf{16} & \to (\bf{4}, \bf{2}) \oplus (\bar{\bf{4}}, \bf{2}) \\
\Psi & \to \psi_{I \alpha} \oplus \tilde{\psi}^{J \beta}
\end{align}
where $I$ is the fundamental index for $\SU(4)$ and $\alpha$ is a
2-component spinor (fundamental)  index for $\SU(2)$.
The reality condition on $\Psi$ implies that $\tilde{\psi}$ is not
independent but rather can be written
in terms of $\psi$, $(\psi^\dagger)^{I \alpha} = \tilde{\psi}^{I \alpha}$.  
Then introducing the notation $\tg^a_{IJ}$ for the Clebsch-Gordon
coefficients $\bar{\bf{4}} \otimes \bar{\bf{4}} \to \bf{6}$, normalized so
that
\begin{equation}
\tg^a (\tg^b)^\dagger + \tg^b (\tg^a)^\dagger = 2 \delta^{ab},
\end{equation}
we can rewrite the above Hamiltonian
\begin{eqnarray} 
\label{h-base} H_0 &=& R \Tr \left( \half \Pi_A^2 - \qtr [X^A, X^B]^2 + \psi^{\dagger I 
\alpha} \sigma^{i \beta}_\alpha [X^i, \psi_{I \beta}] 
\right.\\
\,&-& \left.\half \epsilon_{\alpha \beta} \psi^{\dagger \alpha I} 
\tg^a_{IJ} 
[X^a, 
\psi^{\dagger \beta J}] + \half \epsilon^{\alpha \beta} \psi_{\alpha I} 
(\tg^\dagger)^{a I J} [X^a, \psi_{\beta J}] \right),\\
\label{h-mu} H_\mu &=& \frac{R}{2} \Tr \left( \left(\frac{\mu}{3R}\right)^2 (X^i)^2 + 
\left(\frac{\mu}{6 R}\right)^2 (X^a)^2 +  \frac{\mu}{2 R} \psi^{\dagger I \alpha} \psi_{I \alpha} + i \frac{2 \mu}{3 R} \epsilon^{ijk} X^i X^j X^k \right).
\end{eqnarray}
For later use we also record the kinetic part of the Lagrangian,
\begin{equation} \label{lkin}
\LL_{kin} = \Tr \left(\frac{1}{2R} (D_0 X^A)^2 + i \Psi^{\dagger I \alpha} D_0 
\Psi_{I \alpha} \right),
\end{equation}
which implies the equal time commutators
\begin{align}
[(X^{A})_{mn}, (\Pi_{B})_{kl}] &= i \delta_{ml} \delta_{nk} \delta_{AB}, \\
[(\psi_{I \alpha})_{mn}, (\psi^{\dagger J \beta})_{kl}] &= \delta_{ml}\delta_{nk}\delta^{\beta}_\alpha \delta^I_J,
\end{align}
with
\begin{equation}
\Pi_A = \frac{1}{R} D_0 X^A.
\end{equation}
The lightlike decompactification limit, in which one expects to recover M-theory
on the plane wave \eqref{planewave}, is $N\to\infty$ with $p^+=N/R$ 
held fixed.  Unlike in the BFSS model where only the states with energies
of order $R/N$ survived the large $N$ limit (to keep $p^+ p^-$ of order one in 
Planck units), here we should keep states whose energies
are finite in units of $\mu$ as $N \to \infty$, so that $p^+ p^-$ scales like the 
boost-invariant combination $\mu p^+$.

The BMN matrix model has several nice properties which were 
absent in the original BFSS model.
For example, in the $\mu \to \infty$ limit the Hamiltonian becomes quadratic, which allows one
to calculate physical observables in a perturbation expansion in powers of 
$R / \mu$ 
\cite{matrixperturb}.  In fact, the representation theory of $\SU(4|2)$ makes it
possible to argue that some states have protected 
energies; in particular the energy remains finite as we 
move away from $\mu = \infty$ and these states therefore can be identified
even at finite $\mu$ 
\cite{protected} (see also \cite{kimplefka}).  
Also, the old puzzle of finding transverse 
fivebranes 
in matrix theory \cite{branesfrommatrices}
seems to be more tractable once we include the mass deformation: 
transverse spherical fivebranes arise from strongly coupled dynamics in a 
vacuum containing a large number of coincident ``small'' membranes
\cite{tfivebrane}.

\section{Orientifolding the BMN model}

M-theory on the plane wave \eqref{planewave} possesses a $\IZ_2$ symmetry
which at the supergravity level is just $x^3 \leftrightarrow -x^3, C
\leftrightarrow -C$. (We could also have chosen $x^1$ or $x^2$, but any
other $x^a$ would not work because we would not obtain a symmetry of the
field strength $F_4$.)  So we can consider orientifolding this background and
study M-theory on the resulting geometry. It is known from anomaly
cancellation that in order to define $11$-dimensional supergravity on a
manifold with boundary one has to introduce an extra $E_8$ gauge theory
living only on the boundary \cite{howittenone,howittentwo}.  The same
arguments apply in the presence of the plane wave deformation, so we
expect that M-theory on our orientifolded plane wave will also contain an
extra $E_8$ super Yang-Mills theory living at $X^3=0$.  In the matrix
model we will see this $E_8$ as a global symmetry of the $\mu \to \infty$,
$N\to\infty$ spectrum.

By making the appropriate projection in the BMN
matrix model, and adding extra fermionic degrees of freedom (0-8 strings)
which reflect the coupling of the D0-branes to the gauge theory on the boundary---or, 
equivalently, are necessary to cancel the anomaly 
in the open membrane worldvolume theory \cite{howittenone}---we
obtain a DLCQ description of M-theory on the orbifolded plane wave
which is the Penrose limit of $AdS_7 \times (S^4 / \IZ_2)$
or\footnote{Note that this $\IZ_2$ is acting as an orientifold, so that
$AdS_7\times (S^4 / \IZ_2)$ has $\SO(6,2)\times \SO(4)$ symmetry
\cite{Gimon}.} $(AdS_4)/ \IZ_2 \times S^7$.
The construction is very similar to previous work in the flat space case
\cite{danielsson,motlquat,kimrey,bss,banksmotl,lowehet,reyhet,horavamatrix}.
We now describe it in more detail.

\subsection{The $\IZ_2$ projection}

From now on we single out the index $3$, so $i = 1,2$; $a = 4, \dots, 9$; 
and $A = 1,2, 4,5,6,7,8, 9$.
Then the BMN Hamiltonian \eqref{h-base}, \eqref{h-mu} has a $\IZ_2$ 
symmetry,
\begin{align}
X^3 & \to -(X^3)^T,\label{zetwoA} \\
X^A & \to +(X^A)^T,\label{zetwoB} \\
\psi_{I\alpha} & \to +(\sigma^3)_\alpha^{\ \beta} (\psi_{I\beta})^T
\label{zetwoC}
\end{align}
where $^T$ refers to transposition of the $\U(N)$ indices.
(We could equally well
have used $-\!\sigma^3$ instead of $+\!\sigma^3$ in \eqref{zetwoC} 
above; the choice amounts to a choice
of chirality of the ten-dimensional $E_8$ gauge theory on the boundary.)
Then we can truncate the fundamental fields to their $\IZ_2$ invariant
parts.  For the bosons this gives a real antisymmetric matrix $\frac{i}{2}(X^3-(X^3)^T)$ 
and eight real symmetric
matrices $\half(X^A + (X^A)^T)$, which are denoted $A^3$ and $X^A$ respectively
in the orientifold model.  For the fermions, $\psi_{I\alpha}$ decomposes into 
$\psi_{I+}=\psi_{I1}$ and $\psi_{I-}=\psi_{I2}$ (the notation is meant to 
emphasize the $\sigma^3$ eigenvalue) and the projection \eqref{zetwoC} reduces
these from Hermitian matrices to symmetric and antisymmetric matrices respectively.
When there is 
no confusion we drop $I$, the $\SU(4)$ index. 
The Hamiltonian for these fields can then be
obtained just by truncation of \eqref{h-base} and \eqref{h-mu}; it is
\begin{multline} \label{h-base-z2}
H_0 = R \Tr \left[ \half \Pi_A^2 - \half E_3^2 - \qtr [X^A, X^B]^2 + \half [A^3, X^A]^2 + i \psi_{+}^{\dagger I} [A^3, 
\psi_{I+}] - i \psi_{a}^{\dagger I} [A^3, \psi_{I-}] \right. \\ 
+ \left. \psi_+^{\dagger I} [X^1 - iX^2, \psi_{I-}] + \psi_-^{\dagger I} 
[X^1 + i X^2, \psi_{I+}] + \left( \tg^a_{IJ} X^a \{\psi^{\dagger I}_+, \psi^{\dagger J}_- \} + h.c. \right) \right]
\end{multline}
and
\begin{eqnarray} \label{h-mu-z2}
H_\mu &=& \frac{R}{2} \Tr \left[ -\left( \frac{\mu}{3 R} \right)^2 (A^3)^2 + 
\left( \frac{\mu}{3 R} \right)^2 (X^i)^2 + \left(\frac{\mu}{6 R}\right)^2 
(X^a)^2 \right. \\
\,&-&\left.\frac{2 \mu}{R} A^3 [X^1, X^2] + \frac{\mu}{2 R} (\psi^{\dagger 
I}_+ 
\psi_{I+} + \psi^{\dagger I}_- \psi_{I-}) \right],
\end{eqnarray}
where $E_3 = \frac{1}{R} D_0 A^3$.
Similarly truncating the kinetic Lagrangian \eqref{lkin} gives
\begin{equation}
\LL_{kin} = \Tr \left[ \frac{1}{2R} (D_0 X^A)^2 - \frac{1}{2R} (D_0 A^3)^2 + i 
\psi^{\dagger I}_+ D_0 \psi_{I+} + i 
\psi^{\dagger I}_- D_0 \psi_{I-} \right]
\end{equation}
which leads to the canonical commutation relations
\begin{align}
[(X^A)_{mn}, (\Pi_B)_{kl}] &= \frac{i}{2}(\delta_{ml}\delta_{nk} + 
\delta_{mk}\delta_{nl})\delta_{AB},\\
[(A^3)_{mn}, (E_3)_{kl}] &= \frac{i}{2}(\delta_{ml}\delta_{nk} - \delta_{mk}\delta_{nl}),\\
\{(\psi_{I+})_{mn}, (\psi^{\dagger J}_+)_{kl} \} &= \frac{1}{2}(\delta_{ml}\delta_{nk}+\delta_{mk}\delta_{nl})\delta_I^J,\\
\{(\psi_{I-})_{mn}, (\psi^{\dagger J}_-)_{kl} \} &= \frac{1}{2}(\delta_{ml}\delta_{nk}-\delta_{mk}\delta_{nl})\delta_I^J.
\end{align}

After the $\IZ_2$ projection the gauge group of the matrix model is 
reduced to $\OO(N)$ and the model describes the plane wave \eqref{planewave}. 
We define 
\begin{equation} \label{pplus}
p^+=N/2R.
\end{equation}
To see that this definition is appropriate in the M-theory limit, note that the physics 
of the $\OO(2k)$ matrix model reduces via the Higgs mechanism to that of the 
$\U(k)$ model far away 
from the domain wall, and \eqref{pplus} agrees with the known $p^+=k/R$ in 
this case.  Another way of understanding this rule is that the total $p^+$ of
an object in the original $\U(N)$ model is now divided among two mirror images.
In any case, the correspondence 
\eqref{pplus} implies that the states appearing in the $\OO(N)$ matrix 
models with $N$ odd correspond to fields in spacetime which are {\it 
antiperiodic} around $x^-$. Such states appear because of a Wilson line
around $x^-$, breaking the gauge group $E_8$ to $\SO(16)$.

\subsection{The symmetry superalgebra $\SU(4|1) \times \U(1)$} \label{symm-superalg}

In Appendix B of \cite{matrixperturb} one can find the explicit operator
realization of the superalgebra of 
the BMN matrix model in terms of matrices, together with
all of the commutation relations.  We now
work out the $\IZ_2$ projection of this superalgebra,
which will be relevant for our model.

First we consider the fermionic generators.
The BMN symmetry superalgebra
is generated by 16 kinematical (nonlinearly realized) and 16 dynamical 
(linearly realized) supercharges, which in our $\SO(3) \times \SO(6)$ notation
are denoted $q_{I\alpha}$ and $Q_{I\alpha}$ respectively.
The $q_{I \alpha}$ involve only the $\U(1)$ part of the 
$\U(N)$ matrices:
\beq
q_{I\alpha}=\frac{1}{\sqrt{R}} \Tr (\psi_{I\alpha}).
\eeq
After the projection (\ref{zetwoC}) $q_{I\alpha}$ is identified 
with $(\sigma^3)_\alpha^\beta q_{I\beta}$, so the $q_{I-}$ component is 
projected out, leaving only $q_{I+}\equiv q_{I}$.
Similarly, 
\begin{align}
Q_{I \alpha} = & \sqrt{R} \Tr \left[ (\Pi^a - i \frac{\mu}{6R}X^a) \tg^a_{IJ} \epsilon_{\alpha\beta} \psi^{\dagger J 
\beta} - (\Pi^i + i \frac{\mu}{3R}X^i)\sigma^{i \beta}_\alpha \psi_{I \beta} \right. \\
& \left. + \half [X^i,X^j] \epsilon^{ijk} \sigma^{k \beta}_\alpha \psi_{I \beta} - \frac{i}{2}[X^a,X^b](\tg^{ab})^J_I 
\psi_{J \alpha} + i[X^i,X^a]\sigma^i\tg^a_{IJ}\epsilon_{\alpha \beta}\psi^{\dagger J \beta} \right],
\end{align}
and under
the $\IZ_2$ projection we see that $Q_{I+}$ is projected
out; the only remaining component is $Q_{I-}\equiv Q_{I}$.  Hence the heterotic
plane wave matrix model has 16 real supercharges.

Next consider the bosonic generators.
In the BMN model the $Q_{I \alpha}$ generate a simple 
superalgebra which was identified
in \cite{protected,kimpark} as $\SU(4|2)$.  The bosonic generators of 
$\SU(2) \times \SU(4) \simeq \SO(3) \times \SO(6)$ are $M_{ij}$ and $M_{ab}$, given by
\beqa\label{Mij}
M_{ij}&=&\Tr 
(X^i\Pi^j-X^j\Pi^i+\half\epsilon^{ijk}\psi^\dagger\sigma^k\psi)\ 
,
\cr
M_{ab}&=&\Tr (X^a\Pi^b-X^b\Pi^a+\half\psi^\dagger\tg^{ab} \psi).
\eeqa
Under the $\IZ_2$ action $M_{13}$ and $M_{23}$ have eigenvalue $-1$,
so they are not symmetries of the reduced theory, whereas $M^{12}$ and 
$M_{ab}$ survive the projection.

We also have the bosonic generators $a_a, a_i$ in the $\U(1)$ sector of 
the BMN model,
which appear in the anticommutator of the $q_{I \alpha}$:
\begin{align}
a_i &= \frac{1}{\sqrt{R}} \Tr \left( \sqrt{\frac{\mu}{6R}} 
X^i + 
\sqrt{\frac{3R}{2\mu}} i \Pi^i \right), \\
a_a &= \frac{1}{\sqrt{R}} \Tr \left( \sqrt{\frac{\mu}{12R}} 
X^a + 
\sqrt{\frac{3R}{\mu}} i \Pi^a \right).
\end{align} 
Under the $\IZ_2$ projection all of these operators survive except for 
$a_3$, which is projected out.

Having catalogued the surviving symmetry generators we can now write out their
algebra, which is the $\IZ_2$ projection of that given in \cite{matrixperturb}.  The
important anticommutators are
\beqa \label{susy-commutator}
\{Q^{\dagger I}, Q_J\}&=&2\delta^I_J H - \frac{i\mu}{6} (\tg^{ab})^I_{\ 
J}M^{ab}+\frac{2\mu}{3}\delta^I_J M^{12}\ \cr
&=&2\delta^I_J {\cal H} - \frac{i\mu}{6} (\tg^{ab})^I_{\ J}M^{ab},
\eeqa
\EPSFIGURE{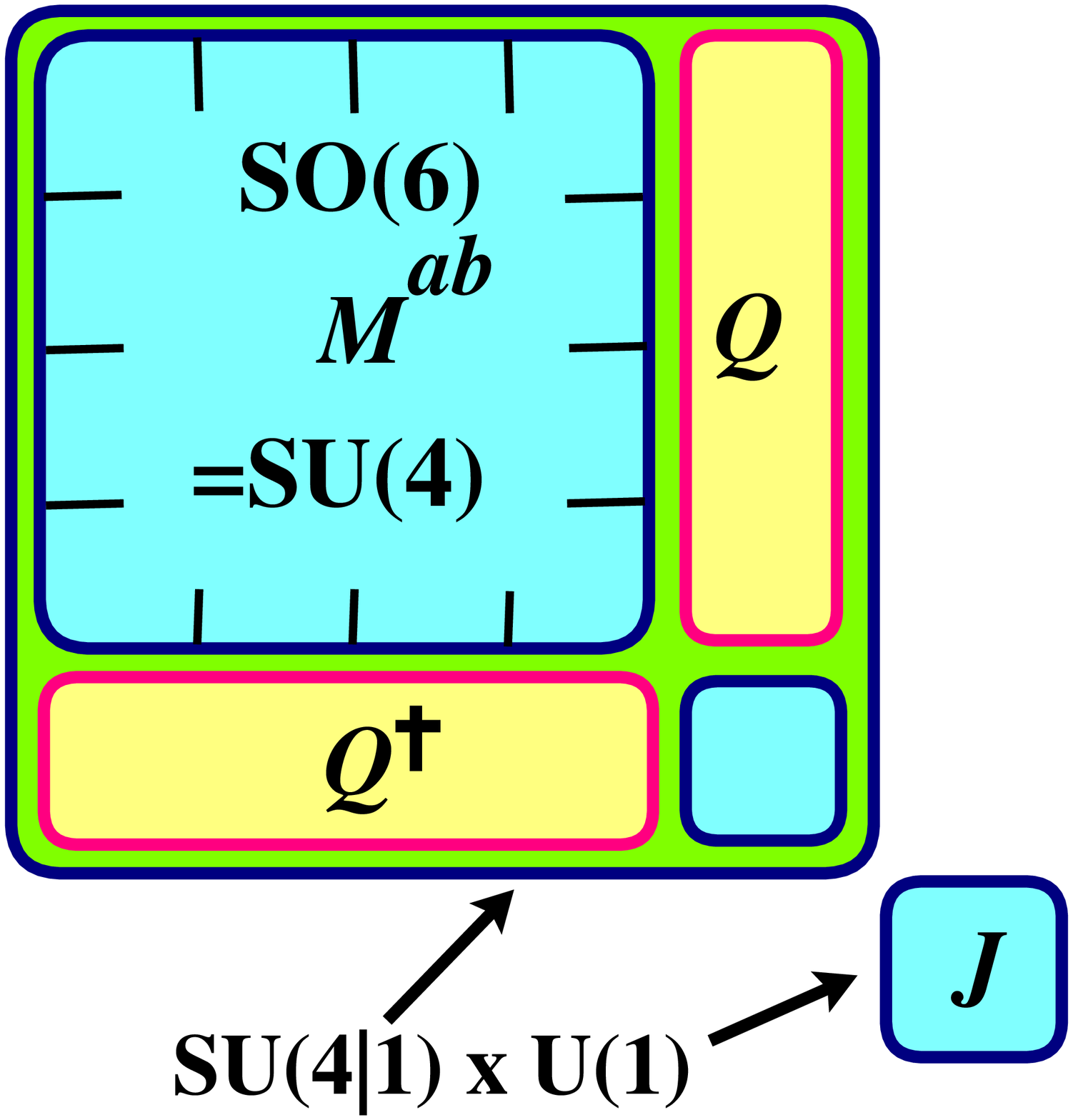,width=63mm}{The structure of $\SU(4|1)\times\U(1)$.
${\cal H}$, not shown, is a diagonal matrix of $\SU(4|1)$, chosen so that
its supertrace vanishes.
\label{supergroupfig}}
\noindent where we have defined
\beq \label{h-def}
{\cal H}= H+\frac{\mu}{3}M^{12},
\eeq
and 
\beqa\label{HQ}
[{\cal H},q_{I}]&=&-\frac{5\mu}{12}q_{I}\ \cr
[{\cal H},Q_{I}]&=&+\frac{\mu}{4}Q_{I}.
\eeqa 

In sum, the bosonic part of the symmetry supergroup in the orientifolded theory is
$\SU(4)\times \U(1)\times \U(1)$, where the two $\U(1)$ factors are 
generated by $H$ and $M^{12}$. However, 
only one combination of the two $\U(1)$'s occurs in the anticommutator
of the supercharges; we have denoted the 
generator of this $\U(1)$ by ${\cal H}$.  This $\U(1)$ combines
with $\SU(4)$ and the dynamical supercharges $Q_I, Q^{I \dagger}$ 
to give the simple supergroup $\SU(4|1)$.
However, there is another combination of the $\U(1)$'s, so the
full symmetry is $\SU(4|1) \times \U(1)$.  The generator of the other 
$\U(1)$, $J$,
must commute with $\SU(4|1)$, so in particular it commutes with $Q_I$.
This requirement implies that we should choose
\beq\label{J-def}
J\,=\,H\!-\!\frac{\mu}{6}M^{12}\,=\,{\cal H}\!-\!\frac{\mu}{2}M^{12}
\,=\,\frac 12(3H\!-\!{\cal H}), 
\eeq
which satisfies
\beq\label{JQ}
[J,Q_I]=0,\ \ \ [J,{\cal H}]=0,
\eeq
and furthermore
\beq
[J,q_I]=-\frac{\mu}{6}q_I.
\eeq
 
So far we have only discussed the superalgebra which acts on gauge invariant (uncharged) states.
More generally we could consider arbitrary states, and in this case the 
commutator \eqref{susy-commutator}
of the supercharges is modified by the addition of a gauge charge on the right side:
\begin{equation} \label{susy-commutator-full}
\{Q^{\dagger I}, Q_J\} = 2\delta^I_J ({\cal H} + A^3_{mn} G_{mn}) - 
\frac{i\mu}{6} (\tg^{ab})^I_{\ J}M^{ab}
\end{equation}
Here $G_{mn}$ represents the charge under the $\SO(N)$ generator $E_{mn}$
which is an antisymmetric matrix: 
\begin{equation}
(E_{mn})_{ij} = \delta_{mi}\delta_{nj} - \delta_{ni}\delta_{mj}.
\end{equation}
The generator $G_{mn}$ can be explicitly written in terms of the fields as
\begin{eqnarray}
G_{mn} &= & \frac 12\left(
E^3_{mk} A^3_{kn} - A^3_{mk} E^3_{kn}\right) -
\frac{1}{2}\sum_A \left(X^A_{mk}\Pi^A_{kn}-\Pi^A_{mk}X^A_{kn}\right)+\\
\,&+&\sum_{I=1}^4\left(\psi_{I+,k[m} \psi^{I\dagger}_{+,n]k}
+\psi_{I-,k[m} \psi^{I\dagger}_{-,n]k} \right)
+i\sum_{r=1}^{16} \lambda^r_{[m}\lambda^r_{n]}.
\end{eqnarray}
Note the $\lambda$ part of $G_{mn}$ which is important and will be 
discussed in the following  subsection.

\subsection{The $\lambda$ fields} \label{lambda-fields}

The field content of the orientifold model is not just the $\IZ_2$ projection
of the BMN model;
we must add 16 real fermions $\lambda^r$ ($r= 1, \dots, 16$) in the fundamental 
representation of $\OO(N)$.  This is precisely analogous to what happens in the
flat space heterotic matrix models
\cite{danielsson,bss}.
In that case one can understand
the need for $16$ fermions in two ways:  if we think of the heterotic matrix
model as describing D0-branes in Type I', then the $\lambda^r$ 
arise from quantization of the 0-8 strings;
on the other hand, if we think of the heterotic matrix model as 
the $\IZ_2$ quotient of the BFSS matrix model, we find a
linear potential for $X$ which destroys translation invariance even far from the domain wall 
unless we add the $\lambda^r$ \cite{bss}.  When we study the spectrum of 
our model we 
will find a similar mechanism which fixes the 
number of $\lambda^r$ to be $16$ in our case as well; namely, 
this number is essential to ensure finiteness of the quantum numbers of
the fuzzy hemispherical membrane's ground state.

Let us consider how the addition of the $\lambda$ fields modifies the superalgebra discussed 
above.  Since the $\lambda^r$ are charged under $\OO(N)$, the Lagrangian in the 
orientifold matrix model contains
a covariant kinetic term 
\begin{equation}
\LL_{kin,\lambda} = i (\lambda^r)^T D_0 \lambda^r.
\end{equation}
However, this cannot be the only term involving
$\lambda$, because $A^0$ transforms nontrivially under the $\IZ_2$ projected supersymmetry 
algebra \cite{bfss}.
This transformation can be compensated by fixing the full $\lambda$ dependent part of the Lagrangian
to be 
\begin{equation} \label{l-lambda}
\LL_\lambda = i (\lambda^r)^T (D_0 - A^3) \lambda^r.
\end{equation}
Then since
$A^0$ and $A^3$ have the same variation under the surviving 16 supersymmetries, 
the full action
will be supersymmetric provided we take the $\lambda^r$ to 
be invariant \cite{bss}.

From \eqref{l-lambda} we find the $\lambda$ dependent 
part of the Hamiltonian (in the gauge $A^0 = 0$):
\begin{equation} \label{h-lambda}
H_\lambda = i (\lambda^r)^T A^3 \lambda^r.
\end{equation}
The full Hamiltonian is
\begin{equation}
H = H_0 + H_\mu + H_\lambda
\end{equation}
with the three pieces given in \eqref{h-base-z2}, \eqref{h-mu-z2}, \eqref{h-lambda}.

The fact that $H$ depends
on the $\lambda^r$ creates a puzzle, since the $\lambda^r$ 
do not appear in the $Q_I$ --- how can this be consistent with 
the commutator
\eqref{susy-commutator}?  This puzzle is resolved (as in the flat space case 
\cite{bss}) by
recalling that the question of whether $\lambda^r$ appear or not in a given generator
is not meaningful if we only consider
the algebra acting on gauge invariant states, because in that case 
we are always free to add gauge charges to the generators.  If we consider the full 
commutator \eqref{susy-commutator-full} acting on arbitrary states, the $\lambda$ 
dependence indeed vanishes
on both sides:  since each $\lambda^r$ is a vector of $\OO(N)$,
the gauge charge $A^3 G$ includes $(\lambda^r)^T A^3 
\lambda^r$, 
which cancels $H_\lambda$ \eqref{h-lambda}.
Everything is therefore consistent provided 
that there is no $\lambda$ dependence in $M^{12}$, so the $\lambda^r$ are 
uncharged under this spatial rotation.  (However, in certain
classical vacua the $\lambda^r$ acquire an effective $M^{12}$---see 
Section \ref{single-membrane-summary}.)

In sum, the generators of $\SU(4|1) \times \U(1)_J$ are given by the naive 
$\IZ_2$ projections of their 
$\SU(4|2)$ counterparts, with the exception that $H$ 
must be modified to include $H_\lambda$ given in \eqref{h-lambda}.

\section{Classical vacua: open fuzzy membranes} \label{classical-vacua}

\EPSFIGURE{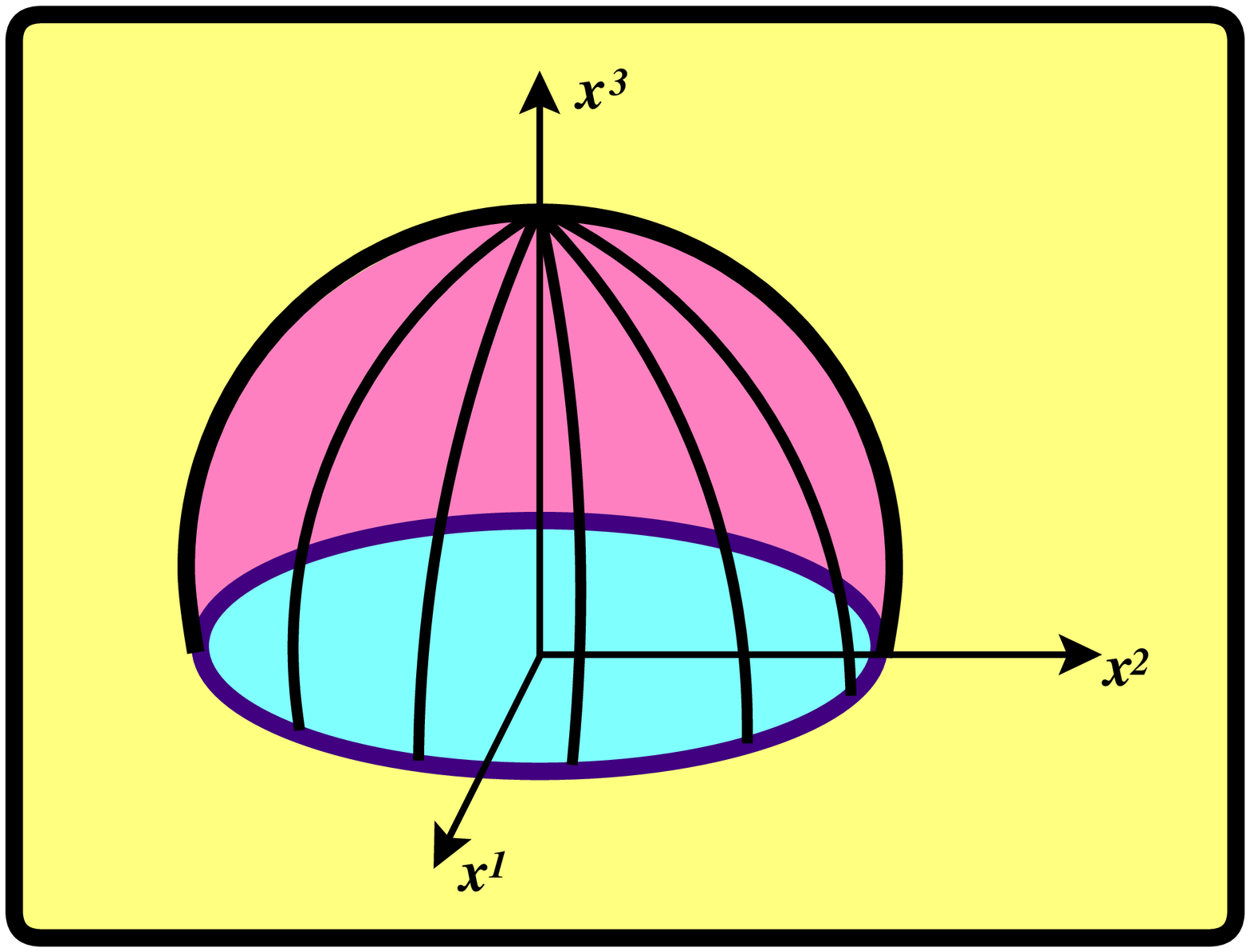,width=83mm}{Hemispherical M2-brane ending
on a Ho\v{r}ava-Witten domain wall $x^3=0$.
\label{hemisphere}}

Any classical solution of the 
\linebreak
original $\U(N)$ BMN model can be described in 
terms of a (generally reducible) $N$-dimensional representation of $\SU(2)$.  Namely,
the three adjoint scalar matrices $X^i$ are the only variables 
that acquire non\-ze\-ro vacuum expectation values, given by 
\cite{{bmn},{matrixperturb}}:
\beq\label{vacua}
X^i = 
\frac{\mu}{3R} J^i,\quad
i=1,2,3
\eeq
where the $c$-number matrices $J^i$ satisfy the standard $\SU(2)$ commutation relations.
Each such representation 
can be decomposed into irreducible representations of $\SU(2)$. 
Then the $X^i$ are block-diagonal and each block,
i.e.\ each irreducible representation of $\SU(2)$, is interpreted as a 
fuzzy spherical 
membrane.  Up to gauge equivalence the solution is uniquely specified
by the dimensions $N_k$ of the irreducible representations, subject to
\begin{equation}
\sum_{k=1}^n N_k = N.
\end{equation}
We denote the solution by $(N_1, N_2, \dots, N_n)$.
The $N_k$ are related to the physical radii of the $n$ fuzzy  
spheres.  For simplicity, consider the irreducible case; then
the ei\-gen\-va\-lues of each $J^i$ belong to the set
$\{-j,-j+1,\dots,j-1,j\}$ with $j=(N-1)/2$, so the physical radius is
\beq \label{membrane-radius}
R_{M2} = \frac{\mu}{3R}\sqrt{j(j+1)} = \frac{\mu}{6R}\sqrt{N^2-1}.
\eeq

Next we want to see whether these classical solutions of the BMN
model give rise to classical solutions of our orientifold model.
Any classical solution of the original model that satisfies the $\IZ_2$ 
constraints (\ref{zetwoA})-(\ref{zetwoC})
becomes a solution of our theory automatically: 
if the action is stationary with respect to {\it any} variations, it is 
of course also stationary with respect to the $\IZ_2$ invariant variations.
Note that in any representation of $\SU(2)$ there is a standard basis
for which two generators are symmetric 
while one is antisymmetric.  Our convention
(which agrees with the one adopted in \cite{matrixperturb}) is that
$J^3$ is antisymmetric while $J^1, J^2$ are
symmetric.
Hence we can identify the antisymmetric 
generator with $A^3$ while\footnote{Our convention disagrees with the standard 
notation used e.g. for Pauli matrices ($j=1/2$) where it is $\sigma^2$, not
$\sigma^3$, that is usually taken to be antisymmetric.} the 
symmetric ones are $X^1,X^2$.
Therefore every solution of the original BMN model
induces a solution of our orientifold model.

Since our model describes a space
in which $x^3$ has been identified with $-x^3$, these solutions are naturally
interpreted in our model as collections of hemispherical 
membranes rather than spherical ones.
This makes sense since it has been known since the early days of matrix 
theory \cite{motlquat,kimrey} that, for large $N$, $\OO(N)$ can approximate the group
of area-preserving diffeomorphisms of the disk, similar to the way $\U(N)$ approximates
the group of area-preserving diffeomorphisms of a closed oriented membrane
\cite{nicolai}.

In summary, the $\OO(N)$ plane wave matrix model inherits the classical solutions of 
its $\U(N)$ predecessor; these are described by $N$-dimensional representations of 
$\SU(2)$, and are interpreted as collections of concentric 
hemispherical membranes, with radii fixed by the dimension of the irreducible 
subrepresentations as in \eqref{membrane-radius}.

\section{Representations of $\SU(4|1)$} \label{su41reps}

As explained in Section \ref{symm-superalg}, the symmetry algebra of
the heterotic BMN matrix model is $\SU(4|1)\times \U(1)_J\times G$ 
(where $G$ is the global symmetry which acts on the $\lambda$ fields;
$G$ manifestly includes $\SO(16)$ and we will eventually see that it
is enhanced to $E_8$ in the large $N$ limit at least in some vacua.)  
In this section we will focus on the \su41 superalgebra
and study its unitary irreducible representations. Our 
discussion in this section is essentially a review of the work by Kac 
\cite{Kac} and Bars et al. \cite{Bars1,Bars2}, specialized to the case of 
$\SU(4|1)$.
A closely analogous and significantly
more detailed discussion in the $\SU(4|2)$ case
may be found in \cite{protected,kimpark}.

Any representation of $\SU(4|1)$ may be decomposed under
$\SU(4)$ into a set of irreducible 
representations, each labelled by the eigenvalue of ${\cal H}$. 
Starting with states $|\psi\rangle$ which form some representation of $\SU(4)$,
the other fermionic and bosonic states in the same supermultiplet can be obtained 
by acting with the supercharges $Q^\dagger_I$ or $Q_I$, $I=1,2,3,4$. Explicitly, a
complete basis for the representation of $\SU(4|1)$ is given by
\begin{equation}
(Q^\dagger_1)^{{\bar s}_1}\ (Q^\dagger_2)^{{\bar s}_2}\ (Q^\dagger_3)^{{\bar 
s}_3}\ (Q^\dagger_4)^{{\bar s}_4}
(Q_1)^{s_1}\ (Q_2)^{s_2}\ (Q_3)^{s_3}\ (Q_4)^{s_4}|\psi\rangle
\end{equation}
with ${\bar s}_i,s_i=0,1$.
(Note that since $\{Q_I,Q_J\}=0$, this representation of $\SU(4|1)$ is finite-dimensional, provided
that the original representation of $\SU(4)$ was finite-dimensional.) The 
simplest examples of representations of $\SU(4|1)$ and their decomposition
under $\SU(4)$ are summarized in Figures \ref{abcrepone}, 
\ref{abcreptwo} at the end of this paper.

Just as for the ordinary Lie group 
$\SU(n)$, there are two standard ways of constructing representations
of $\SU(4|1)$: 
the Kac-Dynkin method and the method of superdiagrams.  We now describe
these two in turn.

\subsection{Kac-Dynkin method, typical and atypical representations} 

As discussed by Kac \cite{Kac} (and reviewed in \cite{protected,kimpark}) 
the representation theory of simple Lie 
superalgebras parallels that of simple Lie algebras.  One 
begins by finding a maximal set of commuting bosonic generators of the 
superalgebra; in our case such a set is $\{H_i\},\ i=1,2,3$, and these
generators span a Cartan subalgebra of 
$\SU(4)$. Then the remaining generators, bosonic and fermionic, can be 
chosen to be eigenvectors of $H_i$; the eigenvalues are called roots. In 
the $\SU(4|1)$ case, only four of the positive roots are 
linearly 
independent; three are simple positive roots of $\SU(4)$ and the 
last is fermionic.  The positive roots and the Cartan algebra 
generators form a maximal subalgebra.  In the Dynkin basis the positive and 
negative roots, $E_i$ and $F_i$, can be chosen such that
\begin{equation}
[H_i, H_j]=0, \ \ [H_i, E_j]=\alpha_{ij}E_j,\ \  [H_i, F_j]=-\alpha_{ij}F_j, \ \
[E_i, F_j]=\delta_{ij} H_i  
\end{equation}
and $\alpha_{ij}$, the Cartan matrix, is

\vspace{-4mm}

\EPSFIGURE[r]{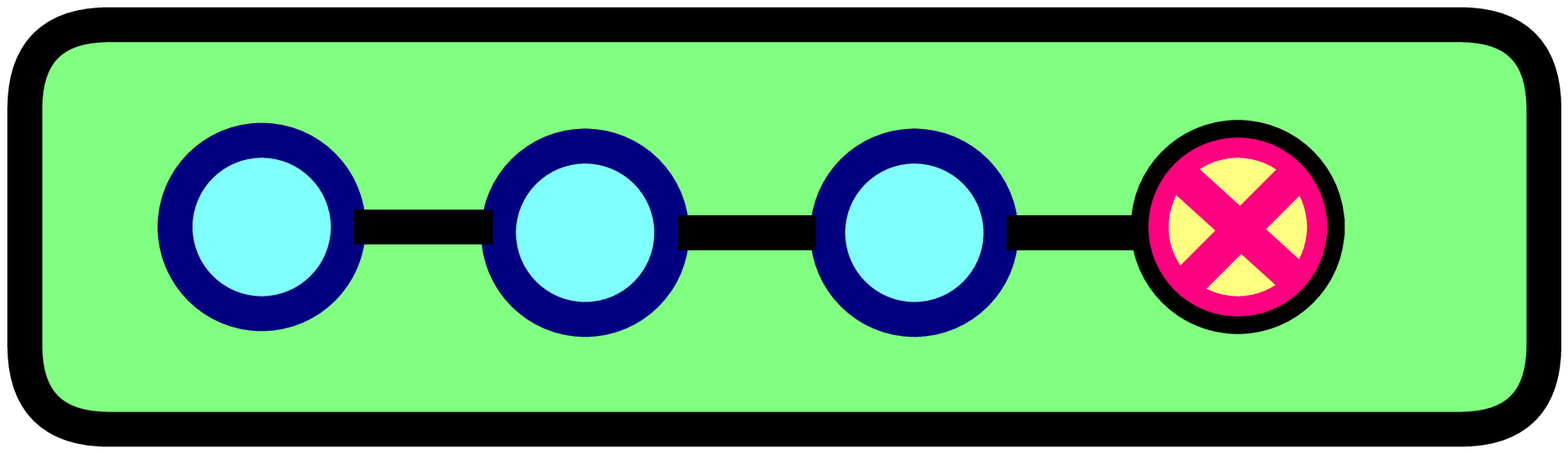,width=55mm}{Kac-Dynkin diagram of
$\SU(4|1)$.\label{kacdynkinfig}}

\begin{equation}
\alpha =\left( \ba{rrrr} 2 & -1 & & \\ -1 & 2 & -1 & \\ & -1 & 2 & -1  \\ 
& 
&-1 & 0 \ea \right).
\end{equation}

This information may be summarized in a Kac-Dynkin diagram with four nodes, where 
the fourth node is fermionic and there are 
$|\alpha_{ij} \alpha_{ji}|$ lines joining  
the $i^{\mathrm{th}}$ and 
$j^{\mathrm{th}}$ nodes; see figure \ref{kacdynkinfig}.
One can then use this basis to construct 
finite dimensional \rep s as one does in the case of ordinary Lie groups. 
Each representation is uniquely determined  by a 
weight vector $\Lambda$, known as the ``highest weight'':
by definition the representation is generated by a single state $|\Lambda\rangle$ 
with weight $\Lambda$ which is annihilated by all the positive 
roots.  In the Kac-Dynkin basis we write
\beq
\Lambda=(a_1,a_2,a_3|a_4),
\eeq
where $(a_1,a_2,a_3)$ are $\SU(4)$ Dynkin labels and hence are 
non-negative integers, while $a_4$ can be any real number. The eigenvalue of 
${\cal H}$, the generator of the diagonal $\U(1) \subset \SU(4|1)$, for this state is
\beq
{\cal H}_0=\mu\left(\frac{1}{12}a_1+\frac{1}{6}a_2+\frac{3}{4}a_3+a_4 \right).
\eeq
The other states in the multiplet can be obtained by the action of the 
$Q_I$'s on the highest weight state:
\beq\label{multiplet}
(Q_1)^{s_1}\ (Q_2)^{s_2}\ (Q_3)^{s_3}\ (Q_4)^{s_4}|\Lambda\rangle
\eeq
where $s_i=0,1$. Note that in principle the multiplet generated in this way may 
be reducible. As argued in \cite{Kac} all the unitary \rep s should have 
positive ${\cal H}$ and unitarizable \rep s should satisfy
\begin{equation}
a_4 \ge 0.
\end{equation}
In general, \ch for a state of the form (\ref{multiplet})
equals ${\cal H}_0+\frac{\mu}{4}\sum_{i} s_i$.  Therefore, 
we can have at most five different \ch levels in the same multiplet,
spaced by $\frac{\mu}{4}$. 
 
If a \rep\ includes all these five different \ch levels it is called {\it 
typical}. For unitary typical \rep s $a_4>0 $ and hence ${\cal H}_0 >
\mu(\frac{1}{12}a_1+\frac{1}{6}a_2+\frac{3}{4}a_3)$.
The dimension of a typical \rep\ $V(\Lambda)$
is simply given in terms of the dimension of the $\SU(4)$ representation
generated by the \hw state:  
\beqa
\!\!\!\!\!\!\!\!\!
\dim(V(\Lambda)) &=& 2^4 \dim (V_0(\Lambda)) \cr
 &=& \frac{2^4}{12} 
(a_1\!+\!1)(a_2\!+\!1)(a_3\!+\!1)(a_1\!+\!a_2\!+\!2)
(a_2\!+\! a_3\!+\!2)(a_1\!+\!a_2\!+\!a_3\!+\!3).
\eeqa

However, it may happen that an irreducible representation includes 
fewer than five \ch levels; in this case the representation is called
{\it atypical}.  As was discussed in \cite{Kac, Bars2}, for 
atypical \rep s we must have
\begin{equation}
a_4\in \{ -\! a_3\!-\!1,\,\, -\! a_3\!-\!a_2\!-\!2, \,\,-\!a_3\!-\!a_2
\!-\!a_1\!-\!1,\,\, 
0\},
\end{equation}
and therefore for the unitary atypical \rep s $a_4$ must vanish.
As a result, the value of ${\cal H}_0$ in an atypical representation is
already determined by the
$\SU(4)$ quantum numbers. In this way atypical 
representations are similar to BPS states; in contrast to the
typical representations, which sit in continuous families parameterized
by $a_4$, an atypical representation is rigid and has no continuous 
deformations.\footnote{Note 
that in contrast to the $\SU(4|2)$ case considered in 
\cite{protected,kimpark}, there are no 
``doubly atypical representations'' in the $\SU(4|1)$ case.}
The dimension of an atypical \rep\ is generally smaller than $2^4 \dim 
(V_0(\Lambda))$.  We will discuss the precise relation to BPS states
in Section \ref{bps}.

\subsection{Young superdiagrams}

\EPSFIGURE{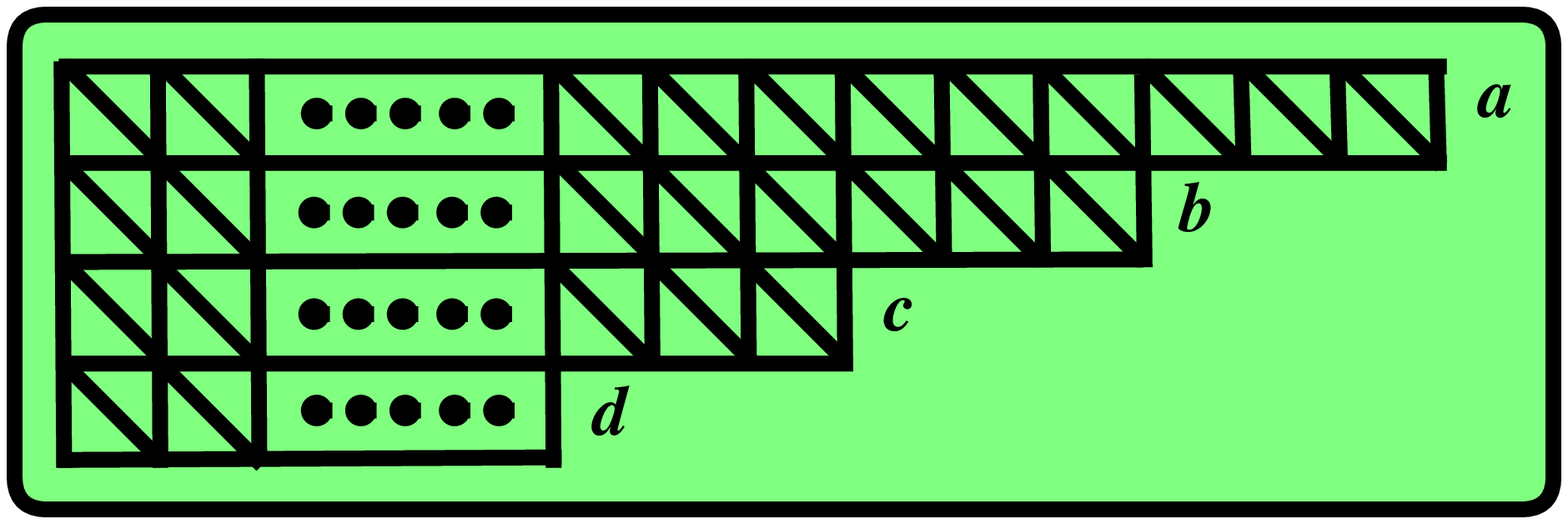,width=85mm}{The most generic $\SU(4|1)$ 
superdiagram.
\label{figureONE}}

In ordinary Lie algebras all 
\linebreak
finite dimensional \rep s can be obtained through 
(tensor) products of finitely many basic irreps. 
\linebreak  
However, due to the existence of 
the continuous parameter $a_4$, this is not true for Lie superalgebras. 
Nevertheless, as we will see in the next sections, all the states in the spectrum 
of the orientifold of the BMN matrix model at $\mu=\infty$ do fit into tensor 
\rep s.\footnote{Note that the spectrum of the matrix model at {\it finite} $\mu$ does not
consist solely of tensor \rep s and hence not all states fit into 
superdiagrams; to see this it is sufficient to note that some states receive perturbative
corrections to $\HH$ \cite{matrixperturb}.}
Furthermore, as shown in \cite{Bars1,Bars2},  for the $\SU(m|n)$ 
case these 
tensor \rep s can be represented by a supersymmetric version of the standard 
Young diagrams.  Here we review, very briefly, some basic facts 
about superdiagrams of $\SU(4|1)$; for a more detailed discussion the 
reader is 
referred to \cite{Bars1,Bars2}. 

To a multiplet with \hw 
$\Lambda=(a-b,b-c,c-d|d)$, where $d$ is a non-negative integer, we associate
the superdiagram depicted in Figure \ref{figureONE}.
The value of \ch for the highest weight state
is simply given by the total number of boxes:
\beq
{\cal H}_0=\frac{\mu}{12}(a+b+c+d).
\eeq

\EPSFIGURE{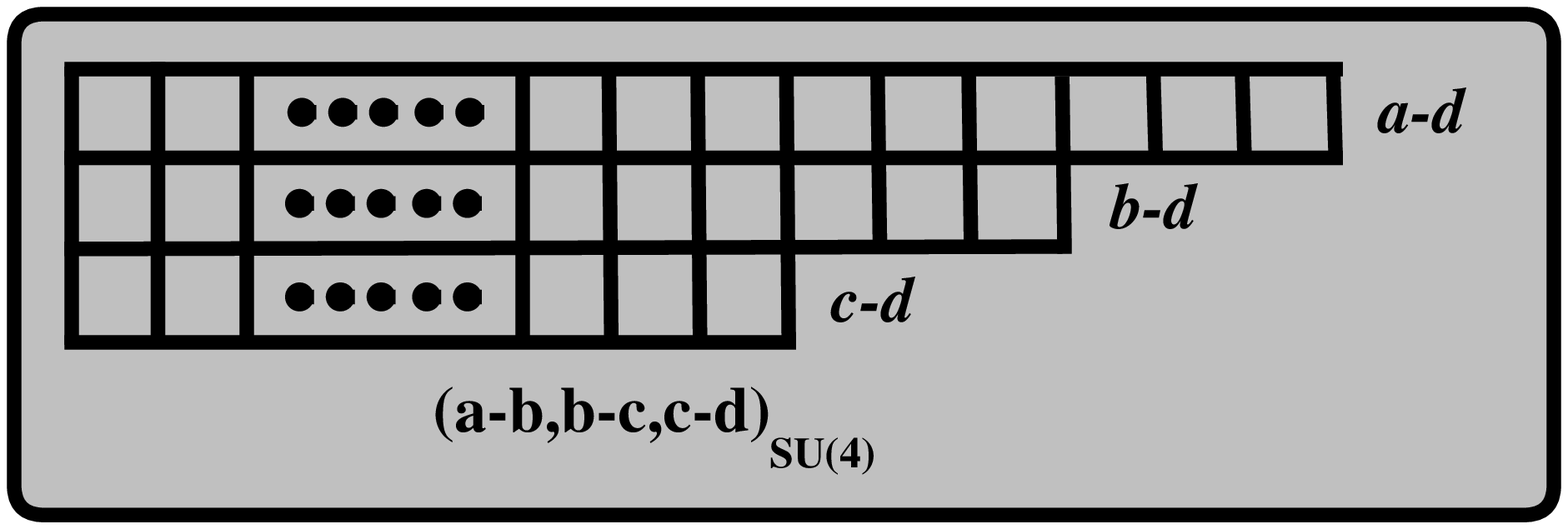,width=85mm}{The $\SU(4)$ \rep\ in which both the states with
the highest ${\cal H}$ states and those with the lowest ${\cal H}$
in the typical multiplet $(a-b,b-c,c-d|d)$ transform.\label{figureTWO}}

The other states in the superdiagram are obtained by acting on the 
\hw 
state (with $\SU(4)$ highest weight $(a-b,b-c,c-d)$) with the supercharges $Q_I$, 
which are in the $(1,0,0)$ \rep\ of $\SU(4)$.

Since we have only four different $Q_I$'s, if $d\neq 0$, the state 
obtained by acting with all of the
$Q$'s on the \hw state has the highest \ch eigenvalue in the 
multiplet, na\-me\-ly ${\cal H}=\frac{\mu}{12}(a+b+c+d)+\mu$.
This state generates a $\SU(4)$ \rep\ isomorphic to that generated by the \hw state
itself; this representation is shown in Figure \ref{figureTWO}. 

\EPSFIGURE{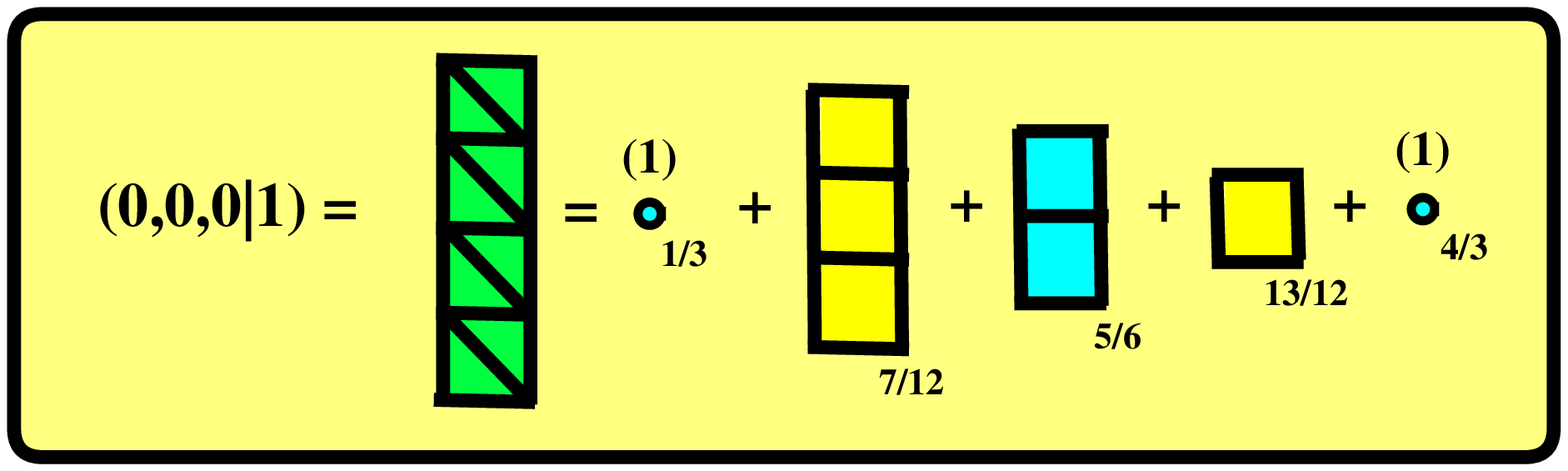,width=85mm}{Expansion of the superdiagram
corresponding to
$(0,0,0|1)$ in terms of $\SU(4)$ Young diagrams. The indices show the
value of \ch (in units of $\mu$) for each state.
\label{figureTHREE}}

To illustrate the expansion of the superdiagram in terms of bosonic and
fermionic modes, as an example we have worked out expansion of
the representation with highest weight 
\linebreak
$(0,0,0|1)$
in Figure \ref{figureTHREE}.

The typical and atypical \rep s can be easily identified in terms of superdiagrams:  
the superdiagrams with four rows ($d\neq 0$) are typical and those with less than four rows 
($d=0$) are atypical. 

Here we summarize some facts about $\SU(4|1)$ superdiagrams:

\begin{itemize}
\item
{\bf i)} Superdiagrams of $\SU(4|1)$ can at most have four rows.
\item
{\bf ii)} The number of ${\cal H}$ levels in the representation
(with steps of $\frac{\mu}{4}$) is the number of rows plus one. 
\item
{\bf iii)} For atypical \rep s the 
value of \ch for the \hw state, and hence for 
all the other states in the multiplet, is completely determined by the 
$\SU(4)$ quantum numbers.
\item
{\bf iv)} The dimension of a typical \rep\ is $2^4$ times the dimension 
of the $\SU(4)$ representation generated from its \hw 
state. The dimension of an atypical \rep\ with \hw 
$\Phi=(a-b,b-c,c|0)$ is
\beqa
\!\!\!\!\!\!\!\!\!\!\!\!
\dim (V(\Phi))= \frac{2^4}{12} &&(a-b+1)(b-c+1)(a-c+2)
\Bigg[\frac{1}{8}(-)^c[2(a-c+2)(b-c+1)-1]+\cr
&&\ \ \ \  +\sum_{l=1}^{c} (-)^{l+1} (a+3-l)(b+2-l)(c+1-l)\Bigg].
\eeqa
\item
{\bf v)} In our conventions 
the \hw state of a superdiagram with an even (odd) number of boxes is bosonic 
(fermionic), in agreement with the spin-statistics relation for $M^{12}$. 
\item
{\bf vi)} Typical \rep s contain equal number of bosonic and fermionic 
states, 
while for atypical ones $B-F\neq 0$. In particular, for the representation with
highest weight $(a-b,b-c,c|0)$,
\beq
B-F=(-)^{a+b+c}\ \frac{1}{2}(a-b+1)(a-c+2)(b-c+1).
\eeq
So $B > F$ for atypical superdiagrams with an even number of boxes, and $B < F$ 
for atypicals with an odd number of boxes.
\item
{\bf vii)} For any multiplet ${\cal X}$,
\beq
\mathop{\mathrm{Str}}_{{\cal X}}({\cal H}) \equiv \sum_{{\cal X}} (-)^F 
{\cal H}= 0. \eeq

\item
{\bf viii)} 
As stated above, we must have $a_4 \ge 0$ for a representation to be unitary,
and $a_4 = 0$ for atypical representations.  If we take a typical representation
$(a-b,b-c,c|d)$ and let $d \to 0$, however, we do not simply get the atypical
representation $(a-b,b-c,c|0)$ but rather the direct sum of two atypicals,
$(a-b,b-c,c|0)$ and $(a-b,b-c,c+1|0)$.  Put another way,
two atypical representations which are of the 
form $(a-b,b-c,c|0)$ and $(a-b,b-c,c+1|0)$ can combine into the typical 
representation $(a-b,b-c,c|\epsilon)$.
The typical representation is created
near $\epsilon=0$, but once it has been created 
the value of $\epsilon>0$---hence also of $\HH$---can 
change continuously while
the $\SU(4)$ content is unchanged.  For example, the typical tensor
representation shown in Figure \ref{figureFOUR} has $\epsilon=1$.

\item
{\bf ix)} The value of \ch in any state of an atypical \rep\ is fixed by the representation's
$\SU(4)$ content, and hence cannot receive corrections {\it unless} there 
exist other atypical representations which can combine with it to form a typical
representation as described above.

\EPSFIGURE{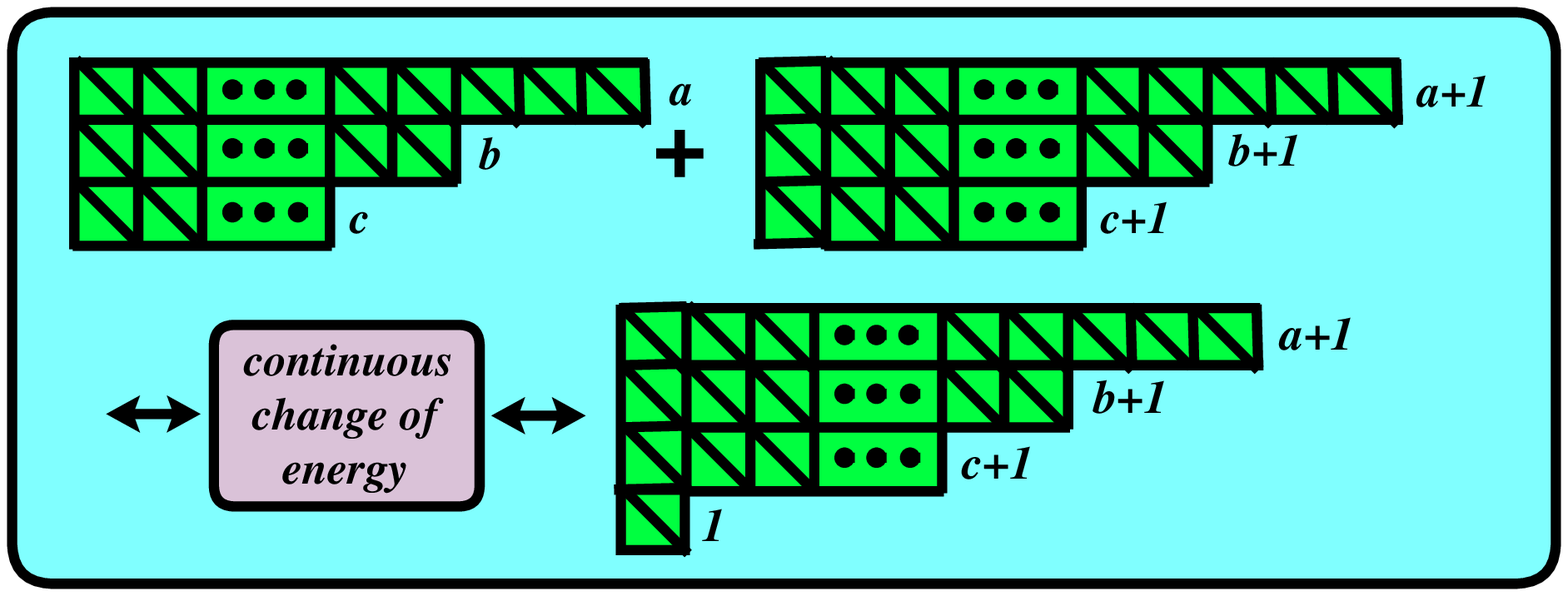,width=140mm}{Two atypical representations which
can combine into the typical representation with highest weight
$(a-b,b-c,c|\epsilon)$ whose $\SU(4)$ content is identical to that of
the tensor representation with heighest weight $(a-b,b-c,c|1)$ shown in 
the figure but whose values of ${\cal H}$ can change continuously.
\label{figureFOUR}}

\item
{\bf x)} Any atypical \rep\ $(a-b,b-c,c|0)$ can be extended to a chain of 
atypical representations, in 
which any representation can combine with either of its two neighbors into 
a typical 
\rep{}. This chain necessarily terminates from one end at $(a-b,b-c,0|0)$. 
Unlike 
the $\SU(4|2)$ case discussed in \cite{protected,kimpark}, in the \su41 case
the chain contains infinitely many diagrams ($c$ can be arbitrarily 
large).\footnote{Although 
$\SU(4|1)$ representation theory does not restrict this chain, in the
finite $N$ matrix model there is a natural cutoff because $c$ is bounded above.
Note that in the matrix model, where we have the external $\U(1)_J$, the 
diagrams
which can combine must also have equal $J$.}  

\end{itemize}

\subsection{BPS states and atypical representations} \label{bps}

As we have discussed, states in atypical representations share some properties with the usual 
BPS states; for example, the values of ${\cal H}$ in such a 
multiplet are completely determined by the $\SU(4)$ quantum numbers. 
One may wonder if these states are also BPS in the usual sense.  Following 
\cite{matrixperturb, protected}, let us check directly
whether any of the states in an atypical superdiagram
is killed by the right-hand-side 
of the superalgebra (\ref{susy-commutator}).  First we rewrite the 
$(\tg^{ab})^I_{\ J}M^{ab}$ term in the superalgebra in terms of three 
$\U(1)$
generators in the Cartan subalgebra of $\SU(4)$; then
\beqa
\!\!\!\!\!\!\!
\langle \psi|\{Q_{I}, Q^{\dagger J}\}|\psi \rangle &\equiv& 
\Delta_{\psi} \delta_I^J \cr
&=&2\left( {\cal 
H}_{\psi}+\frac{\mu}{6}[s_1(I) M^{45}+s_2(I) M^{67}-s_1(I) s_2(I) M^{89}]\right)\delta_I^J,
\eeqa
where $s_i(I)= \pm 1$ are independent functions of the index $I$, $M^{45}, M^{67}, M^{89}$ are the 
eigenvalues of the Cartan generators, and ${\cal H}_{\psi}$ is the eigenvalue of
\ch  for the highest weight state in the 
multiplet. In terms of a Dynkin basis \cite{protected,kimpark},
\beq
M^{45}=\half (H_1+2H_2+H_3),\ \ M^{67}=\half (H_1+H_3),\ \  M^{89}=\half 
(H_3-H_1). 
\eeq
The number of supercharges preserved by the state $|\psi \rangle$ is twice
the number of choices of $s_i$'s for which $\Delta_{\psi}$ vanishes. 

First consider a singlet of $\SU(4|1)$, for which all $M^{ab}$ as 
well as ${\cal H}$ vanish.  Then there are four choices of $s_1, s_2$
for which $\Delta$ vanishes, i.e.\ the vacua preserve all 8 dynamical 
supercharges.  Hence they are \half\ BPS. 
(In the matrix model the vacua will be singlets of $\SU(4|1)$, but they
are not the only singlets.  Note, however, that in the matrix model
we also have $\U(1)_J$, and
$J=0$ for the vacua while $J > 0$ for generic singlets.)

Next let us consider superdiagrams
with only one row.  It is straightforward to 
show that in this case $\Delta_\psi = 0$ precisely when
\begin{equation}
s_1+s_2+s_1s_2=-1.
\end{equation}
In fact this equation is satisfied with three (out of four) choices for 
the $s_i$'s, 
and so the \hw\ states of representations corresponding to
one-row superdiagrams are $\frac{3}{8}$ BPS.

As for the superdiagrams 
with two rows, one can show that in this case $\Delta_\psi = 0$
if and only if $s_1=-1$, so the \hw state of these superdiagrams
is $\frac{1}{4}$ 
BPS. For the diagrams
with three rows, $\Delta_\psi=0$ only if 
$s_1=s_2=-1$; hence the \hw state of such multiplets is $\frac{1}{8}$ BPS.
The diagrams
with four rows are typical representations and do not preserve any supercharges.

In summary, the \hw state of a representation corresponding to a superdiagram
with $r$ rows is
$\frac{1}{8}(4-r)$ BPS (preserves $2\cdot (4-r)$ supercharges). 
We would like to point that this statement applies only to the \hw state of 
the representation. Generically, the other states in the same multiplet (which 
have a higher ${\cal H}$) do not preserve any supercharges.

\subsection{A supersymmetric index} \label{susy-index}

To discuss nonperturbative properties of the spectrum it is useful to 
define an index, similar to \cite{wittenindex}, which is independent of 
the coupling $R / \mu$ (at least provided that no states contributing to
the index become non-normalizable as we vary $\mu$.) 
Since we know the full spectrum of the theory at weak coupling 
($\mu=\infty$) we can evaluate the index and in this way potentially extract some information 
about the non-perturbative spectrum of the matrix model. Our arguments parallel 
those of Appendix A of \cite{tfivebrane}. 

Among the four (complex) dynamical supercharges we single out one, 
$Q^{\dagger}$, which carries charge $\frac{1}{2}$ under three $\U(1)$ 
factors of 
$\SU(4)$:
\beq\label{MQ}
[M^{45},Q^{\dagger}]=[M^{67},Q^{\dagger}]=[M^{89},Q^{\dagger}]=+\frac{1}{2}Q^{\dagger}.
\eeq
Then $Q^{\dagger}$ satisfies 
\beq\label{calE}
\{ Q, Q^{\dagger} \}={\cal H}-\frac{\mu}{6}(M^{45}+M^{67}+M^{89})\equiv {\cal E}
\eeq
From (\ref{HQ}), (\ref{JQ}) and (\ref{MQ}), it is easy to show 
that
\begin{equation}
[Q,{\cal E}]=0,\ \ [J,{\cal E}]=0.
\end{equation}
Furthermore, from (\ref{calE}) we see that the eigenvalues of ${\cal E}$ are all non-negative.

Now we define an index by 
\beq\label{index}
{\cal I}=\mathrm{Str}(e^{ - \beta {\cal E}}).
\eeq
This index receives contributions only from states with ${\cal E}=0$ (which are
necessarily annihilated by $Q$, hence BPS.)  
As usual for an index, ${\cal I}$ gives a lower bound on the number of
states with ${\cal E}=0$.  This bound is actually saturated if
among the BPS multiplets which contribute there 
are none which can combine into typical (non-BPS) multiplets.  We will see in 
section \ref{fivebrane} that in some physically relevant cases we can indeed
show that this bound is saturated.

At first ${\cal I}$ might seem like a rather crude invariant, since
all atypical representations contain BPS states for which 
${\cal E}=0$.  However, there are some other operators which commute with
the supercharges, so we can restrict ${\cal I}$ to their eigenspaces and get
finer information.  
In section 3, we found one such label, $J$. 
Another label which will be convenient later is
\beq
{\cal K}= {\cal H}+2J-\frac {\mu}{2} M^{45}.
\eeq
One can check that both ${\cal K}$ and $J$ have non-negative spectrum.
All classical vacua have ${\cal E}={\cal K}=J=0$.

\section{Spectrum of oscillators about the classical vacua at $\mu=\infty$}

In the limit $\mu \to \infty$ the theory around every
classical vacuum is quadratic and all the degrees of freedom are (bosonic and
fermionic) harmonic oscillators.  We now catalog these oscillators
and their various quantum numbers.  This
discussion lays the groundwork for our later analysis of the physical states
and their implications, which appears in Sections \ref{membrane} and \ref{fivebrane}.
We will first discuss the oscillators about the 
irreducible vacuum and then move on to an arbitrary vacuum.  

\subsection{For the irreducible vacuum} \label{single-membrane-summary}

The spectrum of harmonic oscillators around the irreducible vacuum is summarized in
Table \ref{OscillatorTableOne} below.

\vspace{2mm}

\TABLE{\begin{tabular}{c}
$
\begin{array}{|c|c|c|c|c|c|c|}
\hline 
\mbox{Type}&\mbox{Label}&\mbox{Mass }(H)&\mbox{Spins}&\mbox{Constraint}&
\mbox{Degeneracy}\\
\hline \hline
\SO(6)&x^a_{jm}&\frac j3 +\frac 16&0\leq j\leq N-1&j-m\mbox{ even}&
6\times (j+1)\\
\hline
X^1,X^2,X^3&\alpha_{jm}&\frac j3 +\frac 13&0\leq j\leq N-2&j-m\mbox{ 
even}&
1\times (j+1)\\
\quad&\beta_{jm}&\frac j3 &1\leq j\leq N&j-m\mbox{ even}&
1\times (j+1)\\
\hline
\mbox{Fermions}&\chi^I_{jm}&\frac j3 +\frac 14&\frac 12\leq j\leq N-\frac 
32&j
-m\mbox{ even}&
{\bar 4}\times (j+\frac 12)\\
\quad&\eta_{I,jm}&\frac j3 +\frac 1{12}&\frac 12\leq j\leq 
N-\frac 12&j-m\mbox{ even}&
4\times (j+\frac 12)\\
\hline \hline
\mbox{Gauge}&\gamma_{jm}& 0 &1\leq j\leq N-1&j-m\mbox{ odd 
}&1\times (j+1)\\
\hline \hline
\mbox{Boundary}&\lambda^r_m& -\frac m3&
j=\frac{N-1}2&\lambda^r_{-m}=(\lambda^r_{m})^\dagger
&(16)\times 1 \\
\hline
\end{array} 
$\end{tabular}
\caption{Spectrum of oscillators around the irreducible
vacuum.\label{OscillatorTableOne}} }

In Table \ref{OscillatorTableOne}, the index $m$ always runs from $-j$ to $j$ with step $1$. 
All of the 
harmonic oscillators originating from matrix variables
are complex.
On the other hand, 
$\lambda^r_m$ satisfy a reality condition, so only the oscillators with 
$m<0$ are creation operators.

Now let us explain the individual entries in Table \ref{OscillatorTableOne}.
All of the lines except the one labeled ``boundary'' can be understood as the
$\IZ_2$ quotient of the spectrum in the BMN model.
The full spectrum of the BMN model around this vacuum 
at $\mu=\infty$ has been worked out in Table 1 of
\cite{matrixperturb}; in this limit the theory is just a collection of 
harmonic oscillators, so we only have to check which oscillators 
survive the $\IZ_2$ projection.
For this purpose we need to know the 
behavior of the matrices $Y_{jm}$ under transposition. It is 
straightforward to check that
\beq
Y_{jm}^T=(-1)^{j-m} Y_{jm},
\eeq
which is compatible with what we expect in the continuum ($N\to \infty$) limit, 
since the spherical harmonics satisfy the corresponding constraint
\beq Y_{lm}(\pi-\theta,\phi)=(-1)^{l-m}Y_{lm}(\theta,\phi).
\eeq
Therefore, among the modes $x^a_{jm}$, those with odd $j-m$ are odd under 
the $\IZ_2$ projection; hence out of $2j+1$ states, $j+1$ states survive
in the orientifold model.  Similarly for the $\SO(3)$ fluctuations, which 
give rise 
to the modes written $\alpha_{jm}$ or $\beta_{jm}$ in \cite{matrixperturb}.
The fermionic modes are written $\eta^I_{jm}$ or 
$\chi^I_{jm}$; once again the states with odd $j-m$ are projected out and hence
$4\times \frac{2j+1}{2}$ states survive from each of $\eta^I_{jm}$ and 
$\chi^I_{jm}$.
We have included also the oscillators $\gamma_{jm}$ in the pure gauge 
directions, which must be excited in a specific way 
to preserve gauge invariance, 
and therefore do not contribute to the physical spectrum.  

Finally, to understand the line labeled ``boundary'' in Table \ref{OscillatorTableOne}
recall that in the orientifold
matrix model we have some extra operators which did not occur
in the BMN model, namely the
fermions $\lambda$ in the $({\bf N}, {\bf 16})$ of $\OO(N) \times \OO(16)$, introduced
in Section \ref{lambda-fields}.
The $\lambda$-dependent part of the Hamiltonian, $H_\lambda$, is given by
\eqref{h-lambda}.  If we 
expand $A^3$ about the classical vacuum expectation values (\ref{vacua})
and rescale fields so that 
energy is measured in units of $\mu$, then $H_\lambda$ takes the form
\beq\label{lambda}
\frac{H_\lambda}{\mu}=\frac{1}{3} \lambda^r J^3 \lambda^r + 
\left(\frac{R}{\mu}\right)^{\frac{3}{2}} \lambda^r Y^3 \lambda^r,
\eeq
where $Y^3$ is the fluctuation of $X^3$ about its vacuum value;
in the $\mu\to \infty$ limit the second term, which is an interaction 
with the (bare) coupling $({R}/{\mu})^{3/2}$, drops 
out and only the quadratic part remains important.
In the irreducible vacuum $J^3$
has eigenvalues running from $-\!(N\!-\!1)/2$ to $(N\!-\!1)/2$ (times $\frac{\mu}{3}$).  
From \eqref{lambda}
we see that $J^3$ plays the role of a mass matrix, so there is a natural basis 
in which 
\begin{equation}
[H, \lambda^r_m] = -m \frac{\mu}{3} \lambda^r_m.
\end{equation}
In this basis the $\lambda_m$ satisfy the canonical commutation 
relation
\beq\label{anticommute}
\{\lambda^r_m,\ \lambda^s_n\}=\delta^{rs}\delta_{m,-n}.\ 
\eeq

\subsubsection*{The role of $M^{12}$}

The index $m$ which appears on all of the oscillators in Table \ref{OscillatorTableOne} 
gives the $M^{12}$ eigenvalue.  This fact can essentially be
read off from the expansion of the matrix operators in terms of oscillators, 
given in \cite{matrixperturb}, modulo one small subtlety:  the vacuum 
expectation values \eqref{vacua}  
are not strictly invariant under $M^{12}$, which rotates $X^1$ into $X^2$,
so it does not quite make sense to talk about the $M^{12}$ eigenvalues
of excitations around this classical representative of the
vacuum.  The resolution to this problem is simple:  the 
effect of $M^{12}$ on the vacuum 
can be compensated by adding an $\OO(N)$ gauge transformation with gauge parameter $J^3$,
precisely because in this vacuum 
\begin{equation}
[J^3,X^1] = X^2, \qquad [J^3, X^2] = -X^1.
\end{equation}
The oscillators
then have well-defined quantum numbers under the combined transformation (call it $\M$), 
and from the formulae of \cite{matrixperturb} we can
see that each oscillator has $\M = m$.
Incidentally, this $\M$ occurs naturally in the 
commutator of supercharges;
substituting the vacuum 
expectation value of $A^3$ in \eqref{susy-commutator-full} gives
\begin{equation} \label{susy-commutator-full-vac}
\{Q^{\dagger I}, Q_J\} = 2\delta^I_J (H + \frac{\mu}{3} \M) - \frac{i\mu}{6} (\tg^{ab})^I_{\ J}M^{ab}.
\end{equation}
But at any rate, this discussion is just a convenience which allows us to work with a specific classical
representative of the vacuum for the purpose of determining the quantum numbers;\footnote{Similar
arguments could also be applied to the BMN matrix model.}
in the real Hilbert space one would symmetrize over the full gauge orbit of
\eqref{vacua} to obtain gauge invariant states and oscillators, and then there would be 
no difference between $\M$ and $M^{12}$, so $M^{12} = m$.

Since the $\lambda^r$ did not appear in \cite{matrixperturb} we must argue separately
that they also have $M^{12} = m$.  This we do as follows.  We know that $\lambda^r_m$
carries $H = -m \frac{\mu}{3}$.
On the other hand we have already
seen that the $\lambda^r$ are inert under supersymmetry transformations, so 
the $\lambda^r_m$ commute with $\SU(4|1)$; in particular, they must have $\HH=0$ 
and hence $M^{12} = m$.

\subsection{For general vacua} \label{multiple-membranes-summary}

In a general vacuum $(N_1, N_2, \dots, N_n)$ each individual membrane has
the oscillator spectrum discussed in Section \ref{single-membrane-summary}, but
in addition there are new oscillators arising from the 
off-diagonal blocks $B_{kl}$ that connect two different fuzzy spheres.
While in the 
original model the blocks $B_{kl}$ and $B_{lk}$ were independent, the 
$\IZ_2$ constraint relates these two, leaving only one collection of 
oscillators; the precise constraint is that 
the oscillators which have even (odd) $j-m$ should be symmetric 
(antisymmetric) in the $kl$ indices.
Apart from this constraint the spectrum is identical 
to that in the BMN model, and we can therefore 
copy Table 2 of \cite{matrixperturb}.  (There are no off-diagonal 
components of $\lambda$.)  For convenience it is summarized in Table
\ref{OscillatorTableTwo} below.

\vspace{2mm}

\TABLE{\begin{tabular}{c}$
\begin{array}{|c|c|c|c|c|c|}
\hline
\mbox{Type}&\mbox{Label}&\mbox{Mass }(H)&\mbox{Spins}&
\mbox{Degeneracy}\\
\hline \hline
\SO(6)&x^{a,(kl)}_{jm}&\frac j3 +\frac 16&\frac12 |N_k-N_l|
\leq j\leq \frac12(N_k+N_l)-1&
6\times (2j+1)\\
\hline
X^1,X^2,X^3&\alpha_{jm}^{(kl)}&\frac j3 +\frac 13&\frac12 |N_k-N_l| - 1
\leq j\leq \frac12(N_k+N_l)-2&
1\times (2j+1)\\
\quad&\beta_{jm}^{(kl)}&\frac j3 &\frac12 |N_k-N_l|+1\leq j\leq 
\frac12(N_k+N_l)&
1\times (2j+1)\\
\hline
\mbox{Fermions}&\chi_{jm}^{I,(kl)}&\frac j3 +\frac 14&\frac12 
|N_k-N_l|-\frac 
12\leq j\leq \frac12(N_k+N_l)-\frac
32&
{\bar 4}\times (2j+1)\\
\quad&\eta_{I,jm}^{(kl)}&\frac j3 +\frac 1{12}&\frac12 |N_k-N_l|+\frac 
12\leq j\leq
\frac12(N_k+N_l)-\frac 12&
4\times (2j+1)\\
\hline \hline
\mbox{Gauge}&\gamma_{jm}^{(kl)}& 0 &\frac12 |N_k-N_l|\leq j\leq 
\frac12(N_k+N_l)-1&1\times (2j+1)\\
\hline 
\end{array}$
\end{tabular} \caption{Oscillators in the off-diagonal sector connecting
two fuzzy spheres with sizes $N_l$, $N_k$, in a general 
vacuum.\label{OscillatorTableTwo}} }

\section{The membrane picture} \label{membrane}

By taking an appropriate large $N$ limit in the space of
classical vacua, we can obtain vacua describing
hemispherical classical membranes in M-theory.  The simplest example is the
irreducible vacuum, which consists of a single fuzzy hemisphere at finite $N$;
in the $N \to \infty$ limit we obtain a hemispherical membrane with 
\begin{equation}
R_{M2} = \frac{\mu N}{6 R} = \frac{\mu p^+}{6}.
\end{equation}
More generally we can consider a general vacuum $(N_1, \dots, N_n)$ and
take $N \to \infty$ with all $N_i / N$ fixed; then we obtain $n$ hemispherical 
membranes.

In this section we discuss the physics of these membrane vacua.

\subsection{Zero point energy} \label{zpe}

We begin by computing the energy $H$ of the quantum 
mechanical ground states associated with the classical membrane vacua
in the $\mu \to \infty$ limit.
These zero point energies acquire positive contributions from the bosonic 
oscillators and negative contributions from the fermionic oscillators.
While the bosonic and fermionic contributions always cancelled in the BMN model, the 
situation in our model is a bit more subtle; we will find nonzero 
energies in some cases even for supersymmetric vacua.

We begin with the irreducible vacuum, the large $N$ limit of which
describes a single membrane.
As discussed in Section \ref{single-membrane-summary} the oscillator 
degrees of freedom around the irreducible vacuum are
$x^a_{jm}$, $\alpha_{jm}$, $\beta_{jm}$, $\chi^I_{jm}$, $\eta_{Ijm}$
with $j-m$ even, in addition to the $\lambda^r_m$.  
First let us sum the zero point energies 
over all degrees of freedom except the $\lambda^r_m$.
This gives
\begin{align}
&\frac{\mu}{2} \left( 6 \sum_{j=0}^{N-1} \left(\frac{1}{6} + \frac{j}{3} \right)(j+1) + \sum_{j=0}^{N-2} \left(\frac{1}{3} + \frac{j}{3} \right)(j+1) + \sum_{j=1}^{N} \left(\frac{j}{3}\right)(j+1) \right. \\ 
&\left. - 4 \sum_{j=1/2}^{N-3/2} \left(\frac{1}{4} + \frac{j}{3}\right) \left(j + \half \right) - 4 \sum_{j=1/2}^{N-1/2} \left(\frac{1}{12} + \frac{j}{3}\right) \left(j + \half \right) \right) \\
&=\  \frac{\mu}{3} N^2. 
\end{align}
So far the calculation is identical for odd $N$ and even $N$, 
but for the $\lambda^r_m$ the story is slightly different.  Recall that $m$ 
runs from $-(N-1)/2$ to 
$(N-1)/2$,
$(\lambda^r_m)^\dagger = \lambda^r_{-m}$, and
$\lambda^r_m$ raises $H$ by $\frac{\mu}{3} m$.
The single-membrane vacuum is the state of lowest $H$, 
annihilated by all the $\lambda^r_m$ with $m < 0$; so the zero-point 
contribution
from each of the $\lambda^r$ is
\begin{align}
-\half \sum_{m=1/2}^{(N-1)/2} \frac{\mu}{3} m &= -\frac{\mu}{3} \cdot \frac{N^2}{16}\ \ 
\textrm{\,\,\,\,\,for}\ N\ \textrm{even}, 
\\
-\half \sum_{m=1}^{(N-1)/2} \frac{\mu}{3} m &= -\frac{\mu}{3} 
\cdot\frac{N^2\!-\!1}{16} \  \textrm{for}\ N\ \textrm{odd}.
\end{align}
Hence if we want the $H$ eigenvalue of the single-membrane ground state 
to be independent of $N$, the index $r$ must run from $1$ to $16$; 
in that case the ground state energy is
\begin{eqnarray}
H &=& \,0\mbox{\quad for $N$ even}, \\
H &=& \frac{\mu}{3}\mbox{\quad for $N$ odd}. 
\end{eqnarray}
We could similarly compute the value of $M^{12}$ in the ground state, but 
this direct calculation confirms what 
we already expect:  because of the commutator 
\eqref{susy-commutator}, the fact that the $\SO(6)$-invariant
ground state preserves supersymmetry implies that $\HH = 0$ and hence
\begin{eqnarray}
M^{12} &=&\,0\,\,\, \mbox{\quad for $N$ even}, \\
M^{12} &=&-1\mbox{\quad for $N$ odd.}
\end{eqnarray}
Therefore the ground state is a $\SU(4|1)$ singlet, with 
\begin{eqnarray}
J &=&\,0 \mbox{\quad for $N$ even}, \\
J &=&\frac{\mu}{2}\mbox{\quad for $N$ odd.}
\end{eqnarray}

For a reducible vacuum $(N_1, \dots, N_n)$ 
the oscillator spectrum is slightly more complicated,
as described in Section \ref{multiple-membranes-summary}.  We have
to consider the off-diagonal oscillators, for which the $\IZ_2$ projection relates
$B_{lk}$ to $B_{kl}$ and so reduces the degeneracy uniformly by half relative to 
the original $\U(N)$ model; but in that model the zero point energies always cancel as 
we will show below.
Hence there is no net contribution from the off-diagonal oscillators.  We only have to sum 
over the diagonal ones, which contribute exactly as in the irreducible case:  $\mu / 3$ for each odd $N_i$ and 
$0$ for each even $N_i$.

\subsubsection*{Vanishing of the off-diagonal zero point energy}

It is not too difficult to show that the zero point energy coming from the
off-diagonal degrees of freedom cancels.  Table \ref{OscillatorTableTwo} makes
it clear that the contribution is zero if $N_l=0$: the upper limit for $j$ is always
smaller than the lower limit (exactly by one), and therefore there are no degrees of
freedom at all.  Of course, this conclusion is not surprising because a $N_k\times 0$
rectangle contains no oscillators.  To prove the vanishing of the off-diagonal
zero point energy by mathematical induction, we now consider changing
\begin{equation} N_k\to N_k+1,\qquad N_l\to N_l+1,\label{plusone}
\end{equation}
and check that the contribution to the zero
point energy does not change.  The operation \eqref{plusone} does not change
the lower limits for $j$ while the upper ones $j_{\mathrm{max}}$ get
increased by one.  Setting $j = \frac 12(N_k+N_l)-1$, the angular momentum of the new
multiplet $x^a_{jm}$, the increase of the zero point energy is proportional to
\begin{eqnarray}
\frac{2\Delta H_0}{\mu} &=& 6\times (2j+1)\times \left(\frac 16+\frac j3\right) \nonumber\\
&+&1\times (2j-1)\times \left(\frac 13+\frac{j-1}3\right) \nonumber\\
&+&1\times (2j+3)\times \left(\frac {j+1}3\right)\\
&-&4\times (2j)\qquad\!\!\times \left(\frac 14+\frac {j-1/2}3\right) \nonumber\\
&-&4\times (2j+2)\times \left(\frac 1{12}+\frac {j+1/2}3\right)\quad = \quad 0, \nonumber
\end{eqnarray}
which is the desired result.

\subsection{$E_8$ symmetry} \label{lambda-spectrum}

Here we discuss the effect of the $\lambda$ excitations on the $\mu \to \infty$ 
spectrum.  We will find that in the membrane vacua the $\lambda^r_m$
generate an $E_8$ current algebra.  We begin by considering the one-membrane vacuum.

We want to show that the states
fall into degenerate $E_8$ multiplets and we begin with an illustrative example.
First note that the presence or absence of a $\lambda$ zero mode
is determined by the parity of $N$.
For odd $N$, quantization of the zero modes gives 256 states which are the spin
representation of the Clifford algebra on 16 generators; half of these states are projected 
out by the requirement of gauge invariance under the element $-1 \in \OO(N)$, which acts trivially
on all fields in the adjoint representation but nontrivially on the $\lambda$ fields.
Therefore quantization of the zero modes of $\lambda$ always gives 128 
states in a Weyl spinor representation of $\SO(16)$;
which spinor representation we get depends on whether 
we have an even or odd number of non-zero-mode 
$\lambda$ excitations.  So in particular, there are degenerate ground states 
\begin{equation} \label{gg-odd}
\,\,\,\,\,\,\,\,
\vac_{\bf{128}} \qquad \textrm{($N$ odd)},
\end{equation}
transforming in the 
$\bf{128}$ of $\SO(16)$, with $M^{12} = -1$, $H = {\mu}/{3}$.  

For even $N$ there are 
no $\lambda$ zero modes and the requirement 
from gauge invariance under $-1 \in \OO(N)$ is just that
there be an even number of $\lambda$ excitations.
We can consider the states
\begin{equation} \label{gg-even}
\lambda^r_{-\half} \lambda^s_{-\half} \vac \qquad \textrm{($N$ even)},
\end{equation} 
which transform in the 
adjoint $\bf{120}$ of $\SO(16)$ and have
$M^{12}=-1, H = {\mu}/{3}$.
Note that $M^{12}$ and $H$ match those
of the states we found above for odd $N$.
So combining states from odd and 
even $N$ we can build the $\bf{248}$ of $E_8$, although 
no $E_8$ symmetry is visible at fixed $N$.

Such a separation of the $E_8$ multiplets is not a surprise.  It also occurs in 
flat space heterotic matrix models, in a similar way \cite{ks}; a careful analysis in that case 
demonstrates that these matrix models
at finite $N$ really describe the DLCQ quantization of the heterotic 
string with an $E_8$ Wilson line around the light-like circle, which is responsible 
for the symmetry breaking \cite{sussmotl}.

In fact, one can show more generally in our case that all states built on the membrane vacuum
can be organized into $E_8$ multiplets.  For this purpose it is sufficient to restrict attention to
the $\lambda$ fields, since the
Hilbert space is factorized,
\begin{equation}\label{tensorprod}
\HH = \HH' \otimes \HH_\lambda,
\end{equation}
and $\SO(16)$ acts trivially on the $\HH'$ factor 
(containing all the excitations except $\lambda$).
The factor $\HH_\lambda$ is generated by acting with the $\lambda^r_m$ on the vacuum.
Now the key point is that this Hilbert space is the same as the one obtained by acting 
with 16 fermions $\lambda^r_m$ on the bosonic side of the heterotic 
string, with $H$ in the membrane
picture corresponding to $L_0$ in the string picture (up to the factor $\frac{\mu}{3}$), and 
even/odd $N$ corresponding to 
antiperiodic/periodic boundary conditions around the string.

More precisely, $\HH_\lambda$ at finite
$N$ corresponds to a truncated version of the heterotic string spectrum
and only one boundary condition for the fermions; but there is an obvious 
large $N$ limit of $\HH_\lambda$ where we remove 
the restriction on $m$ in $\lambda^r_m$ and 
also take a direct sum of odd and even $N$ Hilbert spaces.  Then this
large $N$ limit of 
$\HH_\lambda$ carries a natural action of $E_8$, constructed as the zero mode part
of the current algebra just as in the heterotic string.
(Ordinarily in the heterotic string we
consider 32 fermions, but splitting the fermions
into groups of 16 and allowing both periodic and antiperiodic boundary conditions in each group
with independent GSO projections is precisely
what gives the $E_8 \times E_8$ symmetry; here we are focusing just on one group of 16 and so 
we get only a single $E_8$.)  The GSO projection in the heterotic string is replaced in the 
membrane setting by the requirement
of invariance under $-1 \in \OO(N)$ discussed above.  Of course, the isomorphism we have
found here is not a coincidence; in the Ho\v{r}ava-Witten picture of the $E_8 \times E_8$ heterotic
string the fermionic degrees of freedom arise in exactly this way, supported on the boundaries
of cylindrical membranes stretched between the orientifold planes.
 
So we arrive at an attractive interpretation of the modes $\lambda_m^r$:
after the zero-branes have blown up 
into the hemispherical membrane, the $\lambda^r$ become fermionic fields 
propagating  
on the circular boundary, with momentum
modes $\lambda_m^r$ carrying $M^{12} = m$.  The momentum $m$ is 
naturally cut off by the finiteness
of $N$, i.e.\ the fuzziness of the membrane.

If we consider multiple membranes, then each membrane carries
its own $\lambda$ oscillators and its own $E_8$ current algebra.  The
global $E_8$ generators in this case are just the sums of the generators for each individual
membrane.

\subsection{$\SU(4|1)$ multiplets} \label{statesmultiplets}

Next we study the oscillations around the irreducible vacuum
and how they fall into representations of
$\SU(4|1) \times \U(1)_J$.  This representation theory is important because, as
we have seen in Section \ref{su41reps}, the $\SU(4|1)$ quantum numbers
determine whether or not a given state has its energy protected by supersymmetry.

As in the BMN model \cite{protected}, 
the $\SU(4|1)$ symmetry permutes the various
oscillators which generate the physical states acting on the vacuum.  So we begin 
by organizing the oscillators into representations of $\SU(4|1)$; a multi-oscillator
state can then be obtained by taking tensor products.  In contrast to the BMN case
where only atypical representations of $\SU(4|2)$ occurred about the
irreducible vacuum, here we find both
typical and atypical ones.  Namely,
the oscillators listed in Table \ref{OscillatorTableOne} are arranged as follows:  
for each fixed $j$ with $1 \le j \le N$, we find
$j+1$ supermultiplets, generated by the $j+1$ oscillators $\beta_{j m}$ 
with $j-m$ even.  The oscillator $\beta_{j,-j}$ generates a singlet; 
the oscillator $\beta_{j,-(j-2)}$ generates a two-box atypical representation;
and the remaining $j-1$
of the $\beta_{jm}$ generate typical representations.  This representation
content is displayed in Figure \ref{figureELEVEN}.
Note that for any operator ${\mathcal O}$ and fixed $j$,
all the modes ${\mathcal O}_{jm}$ have
the same energy $H$, but their $M^{12}$ vary, so the
\ch and $J$ eigenvalues are different (see \eqref{h-def}, \eqref{J-def}) and
they generally do not sit in the same supermultiplet (in contrast to the situation
in the original BMN model.)

Note that all the superdiagrams corresponding to the $\OO(N)$ matrix theory about the 
irreducible vacuum  
have an even number of boxes.  On the other hand, in any pair of 
atypical representations that can combine into a typical representation,
one should have an odd number of boxes.
Hence {\it all} the atypical multiplets about the single membrane
vacuum in the spectrum of our model 
at $\mu=\infty$ should be perturbatively protected.  We 
stress that this is only true for states about the single membrane vacuum. 
On the other hand, at finite $\mu$ it is quite possible to have 
superdiagrams with an odd
number of boxes, and therefore there may be non-perturbative
shifts in $H$.

\EPSFIGURE{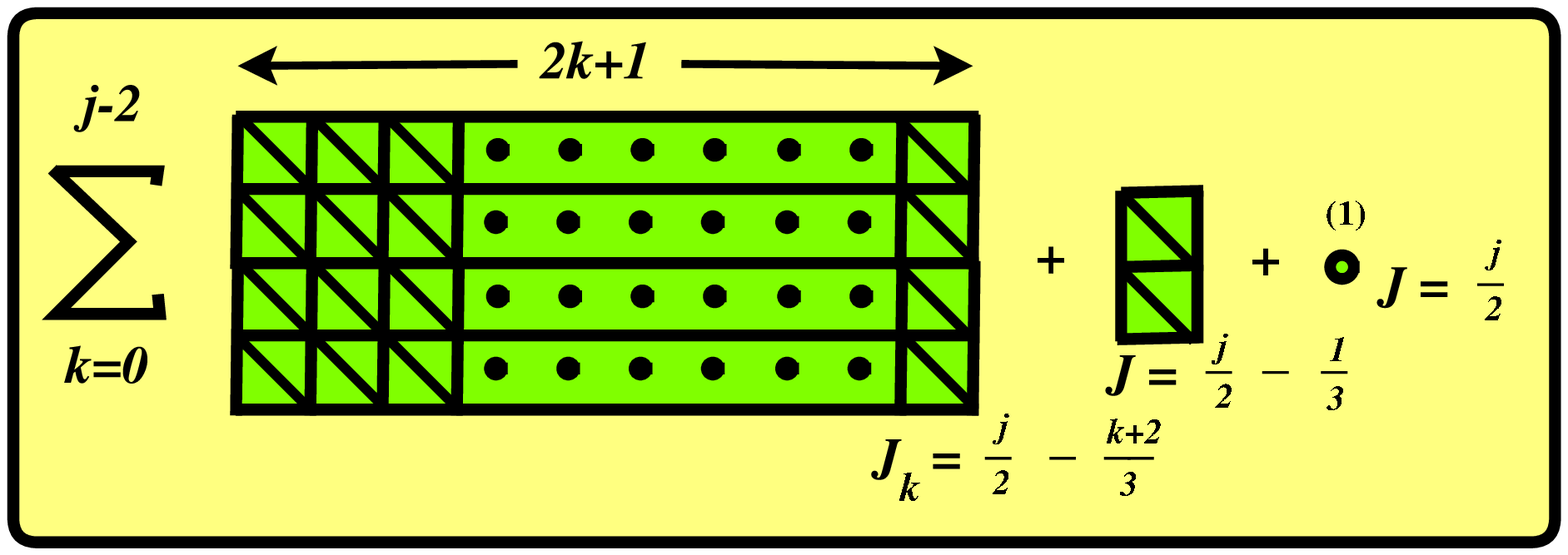,width=140mm}{
Oscillator modes of the orientifold matrix model expanded around the
single-membrane vacuum.
These $j+1$ superdiagrams
correspond to a single-column $2j$-box superdiagram of the $\SU(4|2)$ theory.  
The full spectrum, excluding $\lambda^r_m$, 
is obtained by summing also over $j$ with $1 \le j \le N$.
\label{figureELEVEN}}

A similar analysis could be made for the multimembrane vacua, starting from the
data of Table \ref{OscillatorTableTwo}.
Note however that in the multimembrane vacua we must be careful to account
for a possible residual gauge invariance:  if there are several coincident
membranes ($N_k = N_l$) then there is a subgroup of $\OO(N)$ which leaves
the classical solution \eqref{vacua} invariant, and the oscillators acting
on physical states must be
symmetrized over this subgroup. This guarantees that the indistinguishable 
membranes obey the correct statistics, essentially
because $S_N$ is a subgroup of $\OO(N)$.

\subsection{Giant gluons}

Let us return our attention to the simplest $E_8$ multiplet constructed above
in \eqref{gg-odd}, \eqref{gg-even}.
These states are $\SU(4|1)$ singlets with $J=\frac{\mu}{2}$, that is,
\begin{equation} \label{spins}
M^{12} = -1, \qquad
M_{45} = M_{67} = M_{89} = 0.
\end{equation}
By acting on them with the kinematical supercharges 
we can get other polarizations.  Namely, the operators $q^{I \dagger}$ are a 
spinor of $\SO(6)$ 
and 
raise $M^{12}$ by $\half$; then one can check that acting on 
\eqref{spins}, the $q^{I \dagger}$ produce the quantum numbers 
of the vector multiplet
\begin{equation}
\textbf{8}_v \oplus \textbf{8}_s
\end{equation}
of ${\mathcal N}=1,d=10$ super Yang-Mills, propagating in the plane $x^3 = 0$.

Furthermore, as we saw in the last subsection, all of these vector multiplet
states have energies which are perturbatively protected.  
For the ground state,
which is an $\SU(4|1)$ singlet, this follows from the fact that the 
singlet representation is an atypical
representation.  Then for the other states obtained by acting with $q^{I 
\dagger}$ note that $q^{I \dagger}$
belongs to the decoupled $\U(1) \subset \U(N)$ sector, hence has 
no corrections to
its energy, so the absence of perturbative corrections should
hold just as for the ground state.  Note however 
that the energies
are {\textit{not}} all equal to the $\frac{\mu}{3}$ we found for the 
ground state; rather, by \eqref{HQ} and \eqref{JQ} each $q^{I \dagger}$
increases $H$ by\footnote{Also note that different 
polarizations have different $J$ eigenvalues.} $\frac{\mu}{4}$, so 
the energies range from $\frac{\mu}{3}$ to
$\frac{4\mu}{3}$ according to the rule 
\begin{equation} \label{energy-rule}
H = \mu\left(\frac{5}{6} + \frac{1}{2} M^{12}\right).
\end{equation}

In sum, we have found states of the hemispherical membrane
which transform in the $\bf{248}$ of $E_8$ and have the quantum numbers of the $d=10$ vector
multiplet, with light-cone energies given by \eqref{energy-rule} to all orders perturbatively in 
$R/\mu$; 
we call these states {\bf{giant gluons}}.  Recalling that our matrix model describes DLCQ of
M-theory on the orientifolded plane wave, 
these giant gluon states should be identified with one-gluon states of the gauge field which propagates
in the $X^3 = 0$ plane. 
Similarly, multi-gluon states are identified with states containing several membranes.

\subsection{Giant gravitons}

The appearance of giant gluons here is
analogous to that of giant gravitons in the BMN model; 
in that case there were 16 kinematical supercharges
and their quantization gave the full graviton multiplet through
the ground state degeneracy of the spherical membrane.  In 
our model the graviton multiplet is harder to see
explicitly because we have only 8 kinematical supercharges.
Acting with these supercharges on the even $N$ ground state
generates only 16 states with protected energies, which was
sufficient for the giant gluons but is not sufficient to produce the
whole graviton multiplet.  Some components of the 
$\mu=0$ graviton may indeed have quantum
corrections to their masses at finite $\mu$, and disappear
from the spectrum altogether at $\mu=\infty$.

\subsection{The $\mu \to 0$ limit}

As we have emphasized, one of the advantages of working at finite $\mu$ is that the
problems associated with flat directions in the potential are absent and one can identify
multiparticle states easily.  In particular, we have found states which are candidates for
single- and multiple-gluon states in the DLCQ description of M-theory in the plane 
wave background.
One might ask whether we can use these states to solve the original problem of identifying
the multiple-gluon bound states at $\mu=0$ (or at least to prove their existence.)  In fact, one could
ask a similar question already for the graviton states in the BMN matrix model.  There are
two potential difficulties.  One is that with $16$ supercharges 
our representation-theory arguments are only strong
enough to show that the energies are protected perturbatively
in $R/\mu$.
But even if we assume that we can identify the required states at 
any nonzero $\mu$, with protected energies, there is a more formidable difficulty:
these states may become
non-normalizable in the $\mu \to 0$ limit.
So it seems that our analysis does not allow us to
say anything directly about the existence of bound states in the 
$\mu = 0$ case; the latter problem still requires more subtle analytical tools.

\section{The fivebrane picture} \label{fivebrane}

\subsection{Transverse fivebranes in the BMN model}

It was argued in \cite{tfivebrane} that when the effective coupling about 
a membrane vacuum becomes large there is a dual description in terms of 
transverse fivebranes, which have topology $S^5$ and are extended in
the $X^a$ 
directions. More concretely, consider the membrane vacuum
$(1,1,\dots;2,2,\dots; k,k,\dots,k)$ where the $\SU(2)$ representation
of size $l$ is repeated $M_l$ times (hence $\sum_{l=1}^k lM_l= N$). 
In the 
large $N$ limit when $\frac{M_l}{N}$ is held fixed, the membrane theory 
about this vacuum becomes strongly 
coupled.  However, this large $N$ limit has an alternative
description as $k$ concentric fivebranes, with radii 
\begin{equation} 
\left((R_l)_{M5}\right)^4 = \frac{\mu}{6 R} \sum_{i=l}^k M_i,
\end{equation}
and this dual fivebrane description is weakly coupled and perturbative 
\cite{tfivebrane}. 
In particular, with the above prescription, the trivial $X=0$ vacuum, 
i.e.\ $(1,1,\cdots ,1)$, corresponds to a single 
membrane of radius $R_{M5}^4 = {\mu N}/(6 R)$.
In \cite {tfivebrane} it was checked that the BPS spectrum of 
$k$ non-coincident fivebranes, which is essentially $k$ copies of the
spectrum of geometric fluctuations of a spherical fivebrane plus an 
Abelian self-dual two-form for each fivebrane, 
can be found among the spectrum of exactly 
protected states of the BMN matrix model.\footnote{The 
usual $\SO(5)$ R-symmetry is broken to $\SO(3)$ by
the plane wave background; as discussed in \cite{tfivebrane} this $\SO(3)$ 
symmetry is manifest in the fluctuation spectrum as well as in the matrix model.}

\subsection{Classical fivebranes in the orientifolded plane wave background} 
\label{classical-fivebrane}

Now we want to make a similar analysis in the orientifolded theory.
We begin by working out the classical spectrum which we will try to reproduce
in the matrix model.
First note that the classical $M5$-brane action on the orientifolded
plane wave background in light-cone gauge has a zero-energy configuration corresponding to an
$S^5$ extended in the $x^a$ directions with all $x^i=0$; in particular, $x^3 = 0$,
so the spherical fivebrane is immersed in the Ho\v{r}ava-Witten
domain wall.  
This is a transverse fivebrane since it is extended on the light-cone
time $x^+$ and five spatial directions but not on $x^-$.

Next one can work out the spectrum of geometric 
fluctuations of the spherical fivebrane (these are the ordinary geometric
degrees of freedom obtained from the $\IZ_2$ quotient,
as opposed to the new dynamics arising from 
having the fivebrane embedded in the domain wall.)  Since the calculation is 
essentially similar to that of Appendix B of \cite{tfivebrane} we
do not repeat them here. The result is simply that the modes in the
orientifolded theory are the ones from the parent theory which 
survive the $\IZ_2$ projection.  Among the bosonic excitations of the fivebrane 
spectrum, the fluctuations corresponding to $x^-$, $x^1$, $x^2$ and to the radial direction 
of $S^5$ remain, while $x^3$ and the self-dual two-form are projected out.
Among the fermionic excitations, noting that they are doublets of 
$\SU(2)$, half of them 
which have negative $\sigma^3$ eigenvalue are projected out.
So the 8+8 degrees of freedom of the ${\mathcal N}=(2,0)$ tensor multiplet are
reduced to the 4+4 of the ${\mathcal N}=(1,0)$ hypermultiplet.

\subsection{Single fivebrane vacuum in the heterotic matrix model} 
\label{fivebrane-matrices}

In the BMN model, as discussed in
\cite{tfivebrane}, it was the $X=0$ vacuum---or equivalently the
$(1,1,\dots,1)$ vacuum---that described a single fivebrane.
One might guess that the $X=0$ vacuum also describes
the fivebrane in our model. However, this idea
immediately faces two serious problems:

\begin{itemize}
\item All modes of the $\lambda$ fermions are 
massless around the $X=0$ vacuum
of the $\OO(N)$ theory. By quantizing these $16N$ real fermions, we obtain
a huge representation of $\OO(16)\times \OO(N)$ whose dimension is
$2^{8N}$. In fact, it is
not difficult to see how these states obtained by quantizing fermions
in $({\bf 16},{\bf N})$
decompose under $\OO(16)\times \OO(N)$. One starts with the observation
that the operator
\begin{equation} \label{casimir}
\lambda^r_n\lambda^r_m\lambda^s_m\lambda^s_n
\end{equation}
is related to the quadratic
Casimir operator of $\OO(N)$ because $\lambda^r_{[n}\lambda^r_{m]}$ is proportional
to the generator $G_{mn}$ of $\OO(N)$.  But just by rearranging the fermions
\eqref{casimir} is also related to the quadratic Casimir of $\OO(16)$. A careful counting
of signs and anticommutators reveals that in the more general case
of the group $\OO(M)\times\OO(N)$---in our case $M=16$---we obtain the
identity
\begin{equation}
C^{(2)}_{\OO(M)}(\lambda)+C^{(2)}_{\OO(N)}(\lambda) = \frac{MN}{4} (M+N-2)
\end{equation}
So among the states obtained by quantizing the $\lambda$ fermions, 
smaller representations of $\OO(N)$ are always correlated with larger
representations of $\OO(16)$ and vice versa.
In particular, to obtain physical ground states we would want to look at
the singlets under $\OO(N)$, and these transform as a
huge representation of $\OO(16)$ whose quadratic Casimir increases with $N$. 
It seems that there is no sensible large $N$ limit.

\item A related problem is that the ground state energy of
the $(1,1,\dots,1)$ vacuum in the $\OO(N)$ theory
is $H_0=N\mu/3$.  The reason is simply that
we have a lot of contributions $\mu/3$ from the ground states of
the $N$ one-dimensional fuzzy spheres, since $1$ is odd 
(see Section \ref{zpe}.)  Since $H_0$ becomes infinite as $N \to \infty$,
this means the $X=0$ sector contains no finite energy states in
the large $N$ limit.
\end{itemize}

These problems have a simple resolution. The $(1,1,\dots,1)$ vacuum
actually describes a ``fractional M5-brane''---an object that does not
exist!  The real M5-brane is equivalent to an instanton (not
a half-instanton) in the domain wall gauge theory, and it can always leave
the domain wall. When it does so, its mirror image moves in the opposite
direction. In other words, the fivebrane in heterotic M-theory arises
from a pair of fivebranes in the original theory.  This statement is related to the
well-known fact that the D5-branes in Type~I string theory carry
$\USp(2k)$ symmetry.  Now we see that the relevant vacuum for a single
fivebrane in the heterotic matrix model is 
the $(2,2,\dots,2)$ vacuum. All $\lambda^r_m$ are now massive, and
therefore the vacuum is a singlet under $\OO(16)\times \OO(N)$; 
moreover, since $2$
is even, the ground state energy is $H_0=0$.

\subsection{Spectrum of matrix model about the single fivebrane vacuum} 
\label{fivebrane-spectrum}

\TABULAR[r]{|c|c|c|c|}{\hline
{\rm Mode} &  ${\cal H}$ &  $J$ &   {\rm Degeneracy} \\ \hline
$ \lambda^r_{-\half,k}$ & $0$ & $\frac{1}{4}$ & $16$ \\
\hline
}{The $\lambda$ modes about the $(2,2,\dots,2)$ vacuum. 
\label{osc2-0}}

We would now like to study the excitations of
the $(2,2,\dots,2)$ vacuum, which in the large $N$ limit we want
to identify with a single transverse fivebrane.  We switch notation so that $N$
is the number of $2$-dimensional fuzzy spheres, i.e.\ the gauge group is 
$\OO(2N)$.
The classical solution $(2,2,\dots,2)$ breaks
$\OO(2N) \to \OO(N)$, so all excitations will have to be invariant
under $\OO(N)$. 

\TABULAR[r]{|c|c|c|c|}{\hline
{\rm Mode} &  ${\cal H}$ &  $J$ &   {\rm Degeneracy} \\ \hline
$ x^{\{kl\}}_{00}$ & $\frac{1}{6}$ & $\frac{1}{6}$ & $6$ \\
$\eta^{\{kl\}}_{\half,\half}$ & $\frac{5}{12}$
&$\frac{1}{6}$ & $4$ \\ 
$ \beta^{\{kl\}}_{1,1}$ & $\frac{2}{3}$ & $\frac{1}{6}$ & $1$
\\
\hline
$ \eta^{[kl]}_{\half,-\half}$ & $\frac{1}{12}$  & $\frac{1}{3}$ & $4$ \\
$ \beta^{[kl]}_{1,0}$ & $\frac{1}{3}$  & $\frac{1}{3}$ & $1$ \\
\hline
$ \beta^{\{kl\}}_{1,-1}$ & $0$ & $\frac{1}{2}$ & $1$ \\
\hline
}{The modes from the $\U(1) \subset \U(2)$ part of the BMN model about the 
$(2,2,\dots,2)$ vacuum after a $\IZ_2$ projection. 
\label{osc2-1}}

It is easy to generalize the formalism
of the irreducible $(2)$ vacuum to the $(2,2,\dots,2)$ vacuum: all the oscillators
$x^a_{jm},\alpha_{jm},\beta_{jm},\chi^I_{jm},\eta^I_{jm}$ acquire two extra
indices $k,l=1,2,\dots N$, for the unbroken gauge group $\OO(N)$. Because the $\OO(2N)$
matrix transposition also exchanges the indices $k,l$,
the constraint that
the oscillators with $j-m$ did not exist in the irreducible vacuum
is generalized here to the condition that the
oscillators with $j-m$ odd are antisymmetric in $kl$, which we
indicate by the symbol $[kl]$ in the tables. On the other hand, the oscillators with $j-m$
even are symmetric matrices of $\OO(N)$, which we indicate by the symbol
$\{kl\}$.

Tables \ref{osc2-0}, \ref{osc2-1} and 
\ref{osc2-2} specialize Table \ref{OscillatorTableTwo} to the case
of the $(2,2,\dots,2)$ vacuum.  
(Note that ${\cal H}$ and $J$ are always measured in units of $\mu$.)
In Table \ref{osc2-0} we have listed the $\lambda$ modes.
The $\OO(2N)$ index of $\lambda$ has been decomposed into $2\times N$ as
$\lambda^r_{\pm \half,k}$; hereafter, for simplicity, we suppress the $\pm \half$
and write $\lambda^r_k$ for the creation operator $\lambda^r_{-\half,k}$.
In Table \ref{osc2-1} we have listed the modes from the $\U(1) \subset \U(2)$
part of the BMN model; and in Table \ref{osc2-2} we have listed the remaining 
modes from the BMN model.  

\TABLE{
\begin{tabular}{c}
$
\!
\begin{array}{|c|ccccc|cccc|ccc|cc|c|}
\hline
\mathrm{Mode}& \alpha^{\{kl\}}_{00}\!
&\chi^{\{kl\}}_{\half,\half}\!&
x^{\{kl\}}_{1,1}\!&
\eta^{\{kl\}}_{\frac{3}{2},\frac{3}{2}}\!& \beta^{\{kl\}}_{2,2}\!&
\chi^{[kl]}_{\half,-\half}\!&x^{[kl]}_{1,1}\!&
\eta^{[kl]}_{\frac{3}{2},\frac{1}{2}}\!&\beta^{[kl]}_{2,1}\!&
x^{\{kl\}}_{1,-1}\!&\eta^{\{kl\}}_{\frac{3}{2},-\frac{1}{2}}\!&
\beta^{\{kl\}}_{2,0}\!&
\eta^{[kl]}_{\frac{3}{2},-\frac{3}{2}}\!&
\beta^{[kl]}_{2,-1}\!& \beta^{\{kl\}}_{2,-2}\!\\
\hline
{\cal H}&\frac{1}{3}&\frac{7}{12}&\frac{5}{6}&\frac{13}{12}&\frac 43&
\frac 14&\frac 12&\frac 34&1&
\frac 16&\frac 5{12}&\frac 76&\frac 16&\frac 5{12}&0\\
\hline
J&
\frac 13&\frac 13&\frac 13&\frac 13&\frac 13&
\frac 12&\frac 12&\frac 12&\frac 12&
\frac 23&\frac 23&\frac 23&
\frac 56&\frac 56&1\\
\hline
\mathrm{Degen.}&1&4&6&4&1&4&6&4&1&6&4&1&4&1&1\\
\hline
\end{array}
$\end{tabular}
\caption{The modes
from the $\SU(2) \subset \U(2)$ part of the BMN model about the
$(2,2,\dots,2)$ vacuum which survive the $\IZ_2$ projection.
}\label{osc2-2}}

Now let us discuss the states which may be constructed from
these oscillators.  We write $\OOO^{kl}$ to denote an arbitrary symmetrized product of
oscillators from Tables \ref{osc2-1}, \ref{osc2-2}.  Each such $\OOO^{kl}$ is either
symmetric or antisymmetric in $k,l$ according
to whether it has an even or odd number of antisymmetric constituents.
Then there are three ways to construct
an operator which is an $\OO(N)$ singlet:  we can
take $\Tr \OOO^{\{kl\}}$, $\lambda^r_k \OOO^{\{kl\}} \lambda^s_l$, or
$\lambda^r_k \OOO^{[kl]} \lambda^s_l$, giving respectively ${\bf 1}$, ${\bf 120}$,
or ${\bf 135} \oplus {\bf 1}$ of $\SO(16)$.  We call all of these ``one-oscillator'' states
even though some of them contain three oscillators.  A general state is obtained by acting
with some number of these singlet operators on the vacuum.

\EPSFIGURE{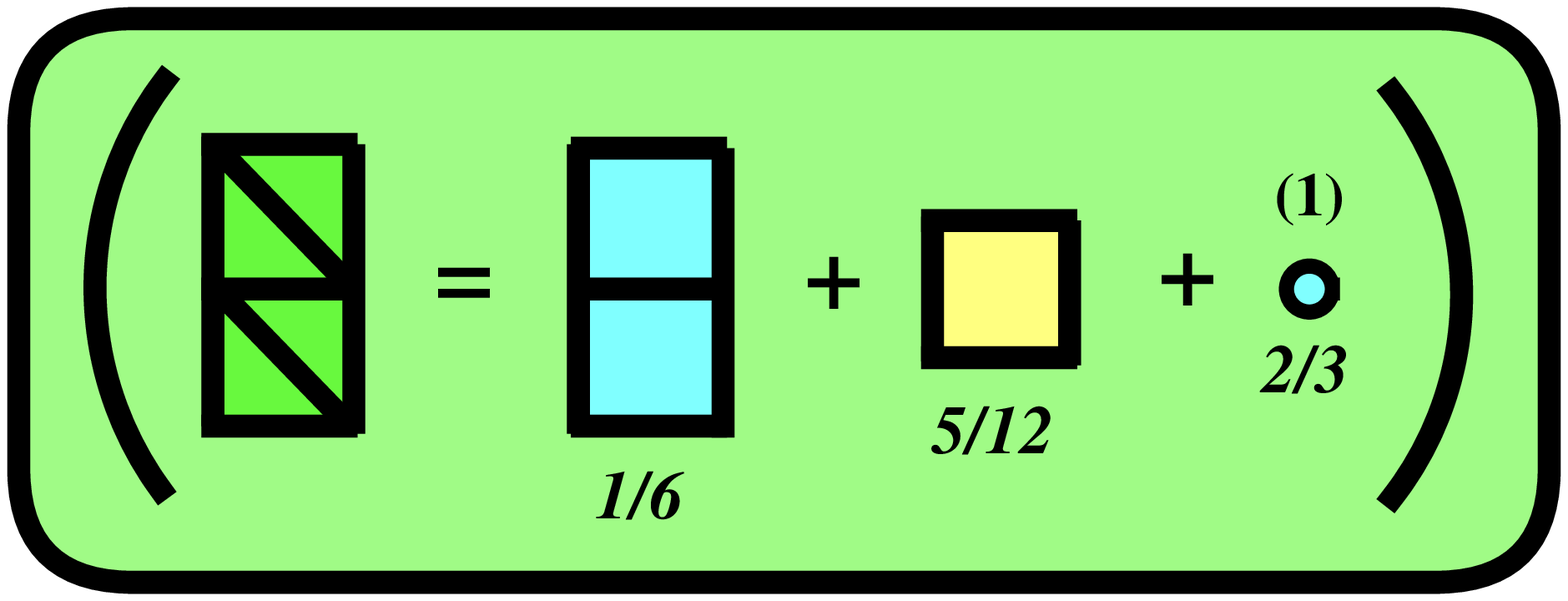,width=64mm}{Two-box multiplet.  This multiplet
appears twice in the spectrum of one-oscillator states
around the $(2,2,\dots,2)$ vacuum.
\label{a2-multiplet}}

To see which of these states are protected, we should analyze the
$\SU(4|1) \times \U(1)_J \times \SO(16)$ multiplets in which they transform.  
However, as we described in Section \ref{su41reps}, the mere fact that a 
particular state is in an atypical multiplet is not enough to ensure that 
it is protected; atypical multiplets whose number of boxes
differ by three as shown in Figure \ref{figureFOUR} can in principle combine
into a typical multiplet and receive corrections.  Of course, in our case where
each superdiagram is also carrying $J$ and $\SO(16)$ quantum numbers, there is the
extra condition that the multiplets which combine
should have equal $J$ and $\SO(16)$ quantum numbers.

\EPSFIGURE{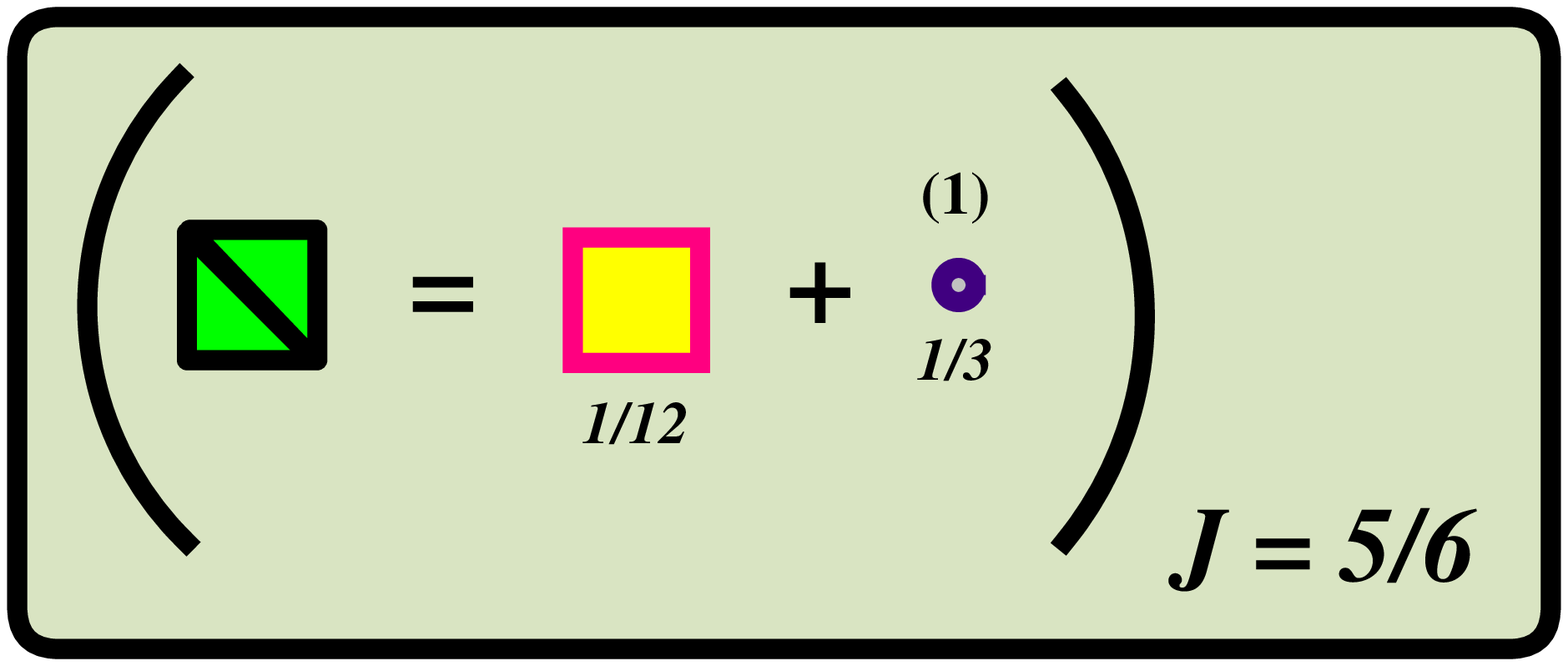,width=64mm}{Single-box multiplet. This multiplet
appears in the spectrum of one-oscillator states around the
$(2,2,\dots,2)$ vacuum.\label{figureTWELVE}}

\subsubsection*{One-oscillator states}

To get a feel for the situation, let us first discuss the one-oscillator states.
First we consider states for which the one oscillator comes from Table 
\ref{osc2-1}.  Their quantum numbers are as follows:

\vspace{2mm}

\noindent${\bullet}$ $\Tr \beta_{1,-1}|0\rangle\ $ is an \su41 singlet with $J=\half$.

\noindent${\bullet}$ $\Tr x_{00}|0\rangle\ $ is the highest weight state of a \su41 
two-box \rep, shown in Figure \ref{a2-multiplet}, with $J=\frac{1}{6}$.

\noindent${\bullet}$
$\lambda^r_k \eta^{[kl]}_{\half,-\half}\lambda^s_l \vac$ and $\lambda^r_k \beta^{[kl]}_{1,0}\lambda^s_l \vac$, both in ${\bf 135 \oplus 1}$ of $\SO(16)$,
form a single box representation of $\SU(4|1)$ with $J=\frac{5}{6}$, shown in
figure \ref{figureTWELVE}.

\noindent${\bullet}$  $\lambda^r_k \beta^{\{kl\}}_{1,-1}\lambda^s_l \vac$
is in the ${\bf 120}$ of $\SO(16)$, $\SU(4|1)$ singlet, with $J=1$.

\noindent${\bullet}$ $\lambda^r_k x^{\{kl\}}_{0,0}\lambda^s_l \vac$ 
is in the ${\bf 120}$ of $\SO(16)$ and is the highest weight state of a $\SU(4|1)$
doublet, like that pictured in Figure \ref{a2-multiplet} except that it has $J=\frac{2}{3}$.

\vspace{2mm}

Next consider the states for which the one oscillator comes from Table \ref{osc2-2}.
These are summarized in Figures \ref{figureSO(16)singlet}, \ref{figureSO(16)other}.

\EPSFIGURE{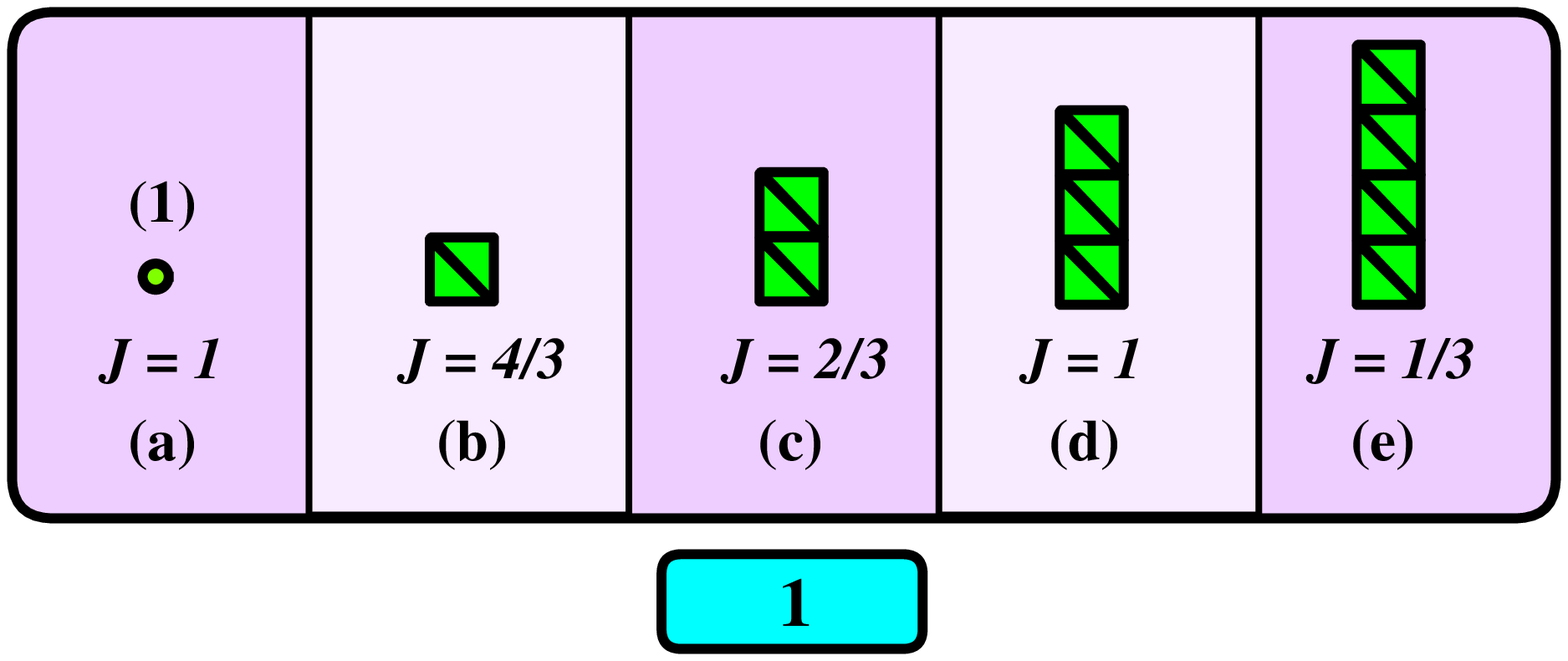,width=120mm}{Multiplets
built from a single oscillator of Table
\ref{osc2-2} which are $\SO(16)$ singlets. The highest weight
states of these multiplets are:
{\bf (a)} $\Tr \beta_{2,-2}|0\rangle$;
{\bf (b)} $\lambda^r \eta_{\frac{3}{2},-\frac{3}{2}}\lambda^r |0\rangle$;
{\bf (c)} $\Tr x_{1,-1}|0\rangle$;
{\bf (d)} $\lambda^r \chi_{\frac{1}{2},-\frac{1}{2}}\lambda^r |0\rangle$;
{\bf (e)} $\Tr \alpha_{0,0}|0\rangle$. Note that except {\bf (e)} all the
multiplets are atypical.\label{figureSO(16)singlet}}

\EPSFIGURE{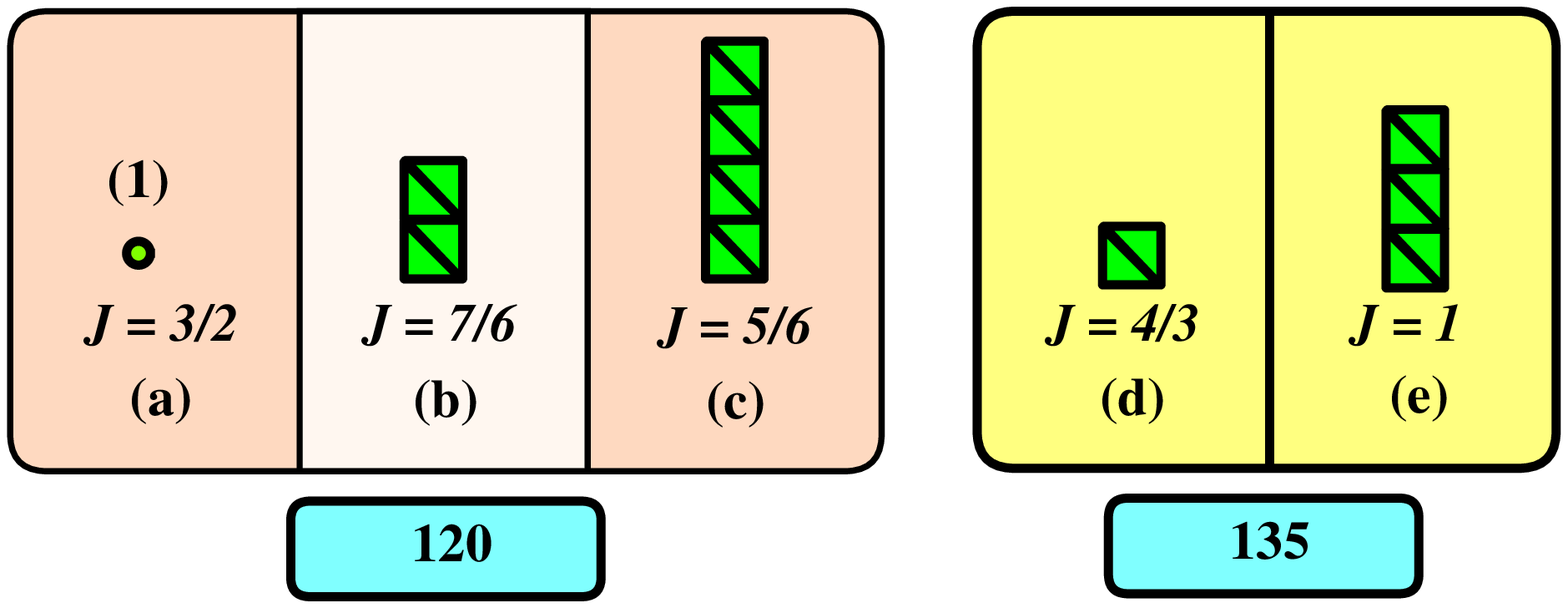,width=145mm}{Multiplets
built from a single oscillator of Table
\ref{osc2-2} which are in ${\bf 120}$ of $\SO(16)$
{\bf (a, b, c)} or ${\bf 135}$ of $\SO(16)$ {\bf (d, e)}.
The highest weight states of these multiplets are:
{\bf (a)} $\lambda^r \beta_{2,-2}\lambda^s |0\rangle$;
{\bf (b)} $\lambda^r x_{1,-1}\lambda^s |0\rangle$;
{\bf (c)} $\lambda^r \alpha_{0,0}\lambda^s |0\rangle$.
{\bf (d)} $\lambda^r \eta_{\frac{3}{2},-\frac{3}{2}}\lambda^s -
\frac{1}{16}\delta^{rs}\lambda^p
\eta_{\frac{3}{2},-\frac{3}{2}}\lambda^p |0\rangle$;
{\bf (e)} $\lambda^r \chi_{\frac{1}{2},-\frac{1}{2}}\lambda^s -
\frac{1}{16}\delta^{rs}\lambda^r \chi_{\frac{1}{2},-\frac{1}{2}}\lambda^r
|0\rangle$.\label{figureSO(16)other}}

Having listed all the one-oscillator states we learn two lessons.
One is that in contrast to the irreducible vacuum, the $(2,2,\dots,2)$ vacuum
can support excitations corresponding to superdiagrams of $\SU(4|1)$ with
an odd number of boxes.  As a result it is possible for diagrams to combine
and the second lesson is that indeed this happens:  for example, in Figure \ref{figureSO(16)singlet}
the multiplets labeled {\bf (a)} and {\bf (d)} can combine to form a typical 
representation.

One can make a similar analysis for two-oscillator states, for which the
representation content is obtained by symmetrized tensor products of the multiplets
listed above.  This analysis is described in Appendix \ref{two-osc-appendix}.

\subsubsection*{Protected states:  geometric fluctuations of the fivebrane}

In general we have seen that it is difficult to argue that an atypical multiplet 
is protected even perturbatively, nonetheless as we will argue some particular 
ones are. Consider the two-row atypical multiplet shown in 
Figure \ref{fig-1}.

\EPSFIGURE{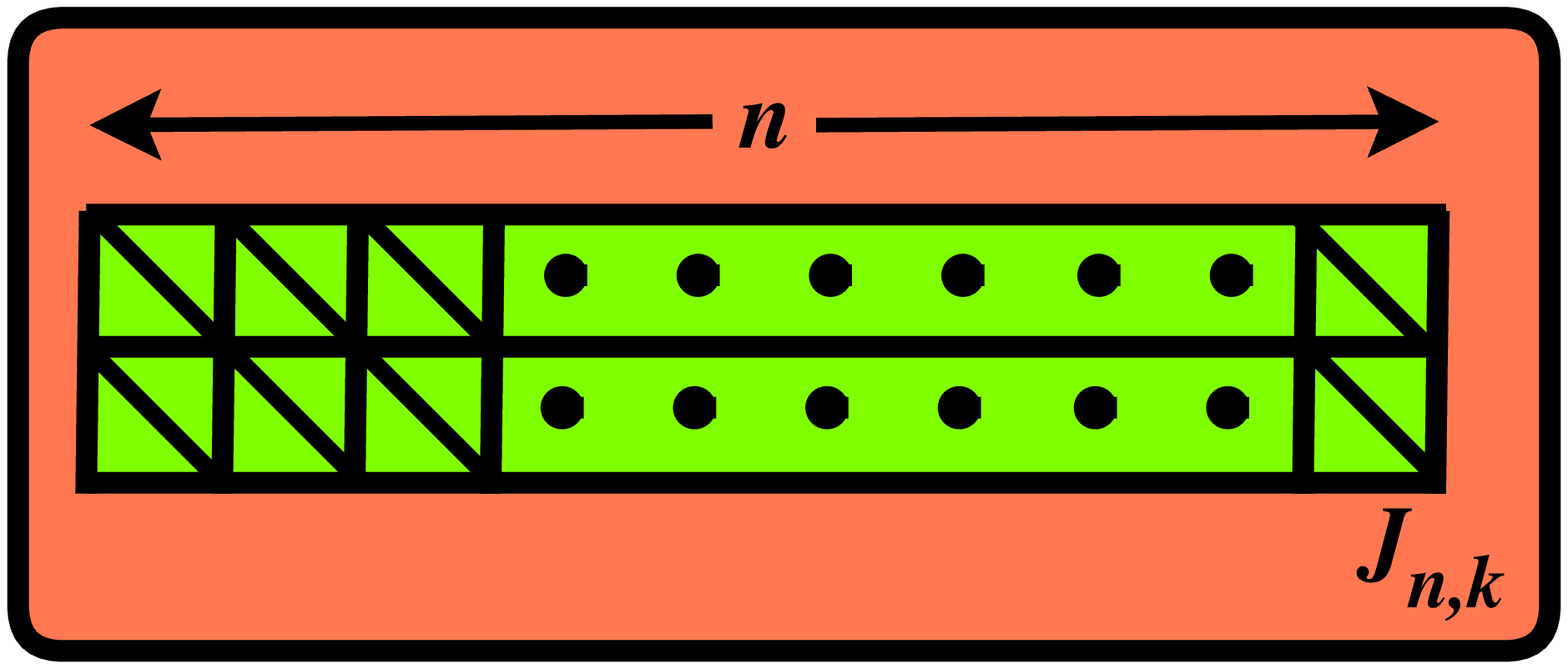,width=85mm}{An atypical multiplet
which occurs in the spectrum around the $(2,2,\dots,2)$ vacuum. \label{fig-1}}

This multiplet occurs many 
\linebreak
times in the spectrum, with lowest $\HH$ states
\begin{equation} \label{two-row-state}
S_{a_1 \cdots a_n} \Tr \left( x_{00}^{a_1} \cdots x_{00}^{a_n} (\beta_{1,-1})^k 
\right) \vac
\end{equation}
where $S_{a_1 \cdots a_n}$ is a totally symmetric traceless tensor of $\SO(6)$.
The state \eqref{two-row-state} is an $\SO(16)$ singlet and has
\begin{equation}
J_{n,k} = \frac{n}{6} + \frac{k}{2}.
\end{equation}
This is an atypical multiplet, but is it really protected?
In fact, there is another multiplet in the spectrum which 
is a candidate to combine with it, namely the one with lowest $\HH$ states
\begin{equation} \label{two-row-partner}
S_{a_1 \cdots a_n} \lambda^r \left( \chi_{\half, -\half} x_{00}^{a_1} \cdots 
x_{00}^{a_n} (\beta_{1,-1})^l \right)\lambda^r  \vac
\end{equation}
This is again an $\SO(16)$ singlet, has $\SU(4|1)$ content given in Figure 
\ref{fig-2}, and
\begin{equation}
J_{n,l} = \frac{n}{6} + \frac{l}{2} + 1.
\end{equation}
\EPSFIGURE{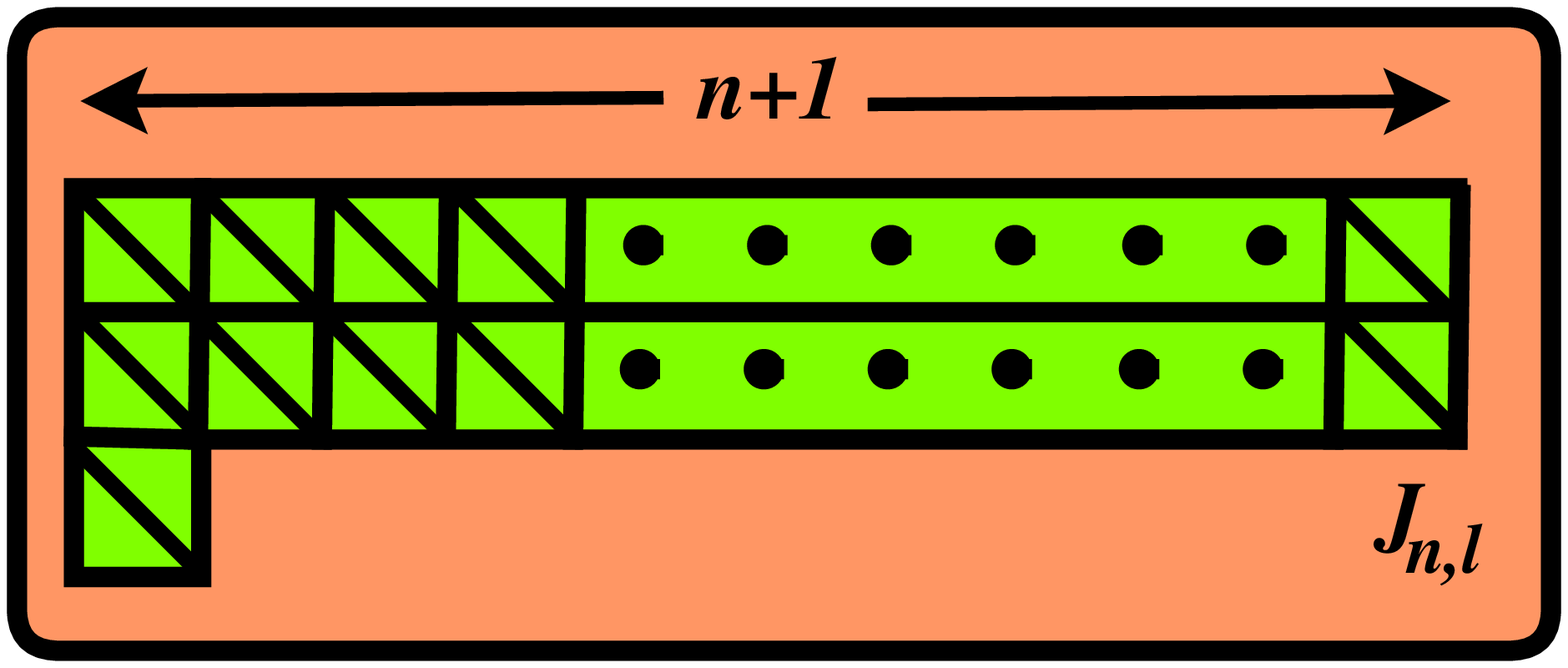,width=85mm}{An atypical multiplet
which could combine with the multiplet in Figure \ref{fig-1}.\label{fig-2}}

So for the two multiplets with highest weights \eqref{two-row-state}, 
\eqref{two-row-partner}
to combine, we must have $k = l + 2$.  
Hence the states \eqref{two-row-state} with $k \ge 2$
are not protected, but the ones with $k = 0,1$ still could be protected.
In fact they have no partners in the $\mu = \infty$ spectrum and hence
their energies cannot receive any corrections in perturbation theory.

\EPSFIGURE{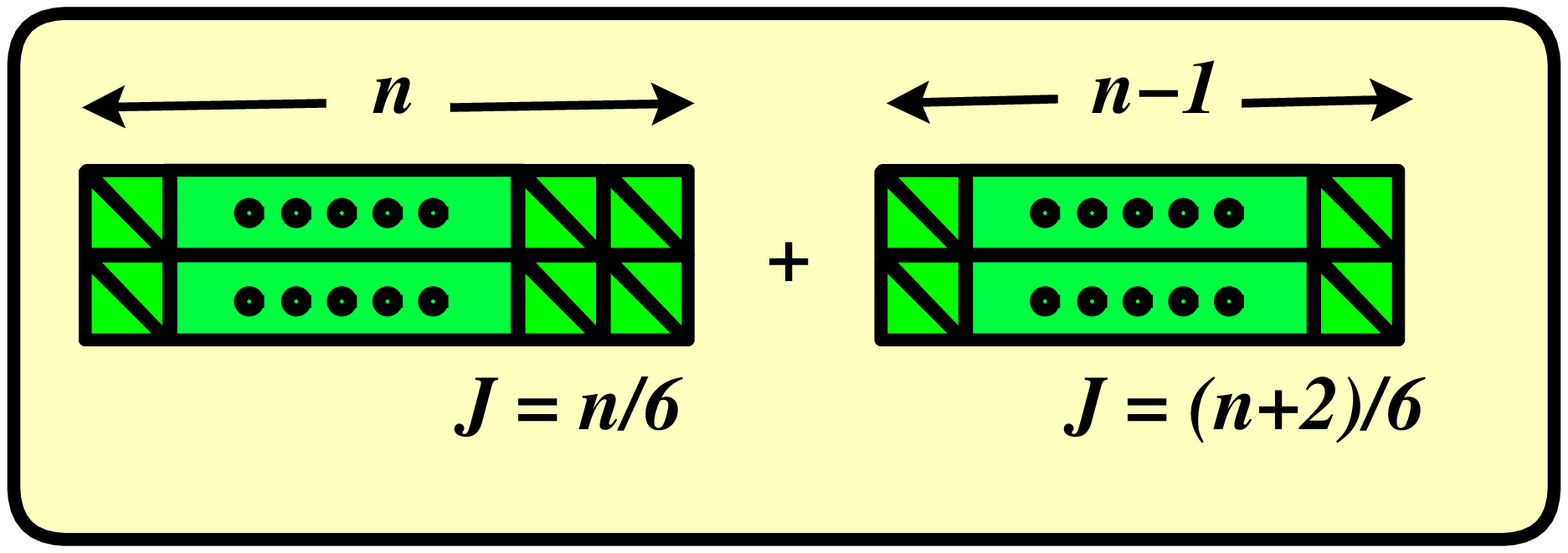,width=120mm}{Two atypical representations which are
exactly protected. The one with $2\times n$ boxes has ${\cal K}=0$ and the other
one ${\cal K}=1$.  These correspond to geometric fluctuations of
the fivebrane.\label{two-protected}}

Note that there are multiplets with the same $\SU(4|1)$ representation
content as shown in Figure \ref{fig-1}, but with a different highest weight
state, e.g.
\begin{equation}
S_{a_1 \cdots a_n} \Tr \left( x_{00}^{a_1} \cdots x_{00}^{a_{n-2}} x_{1,-1}^{a_{n-1}} 
x_{1,-1}^{a_n}\right) \vac .
\end{equation}
However, all such multiplets would have $J > \frac{n+2}{6}$ and hence, as we
argued above, they have potential partners in the spectrum with which they
can combine to form typical multiplets; therefore they are not protected.

So we have shown that the atypicals of the form shown in Figure \ref{two-protected} 
are perturbatively protected, i.e.\ one cannot find any multiplets 
with which they could combine in the spectrum of the matrix model about the 
$(2,2,\cdots, 2)$ vacuum.

In fact, they are protected even nonperturbatively.  To
see this one has to check two things:  first, that the partners of these
states do not appear about any vacuum of the matrix model at $\mu =
\infty$; second, that no typical multiplet appears which could split 
at finite $\mu$ into one of the partners 
of these states.  Both of these can be checked directly by looking at the 
oscillator content of the theory.  For this purpose it is convenient to
exploit the label ${\cal K}$ defined in Section \ref{susy-index}, which we recall:
\begin{equation}
{\cal K} = {\cal H}+2J-\frac{\mu}{2} M^{45}.
\end{equation}
This ${\cal K} \ge 0$ for all oscillators in the theory, and the
multiplets depicted in Figure \ref{fig-1} have 
\begin{equation}
{\cal K}=\frac{\mu n}{6}+2\mu \left(\frac{n}{6}+\frac{k}{2} \right)-\frac{\mu}{2}n= \mu k.
\end{equation}
So $k=0,1$ are relatively hard to make; for these values 
there are not many states carrying ${\cal K} = k \mu$ and one can check directly
that they do not form the dangerous multiplets which could give corrections
to the ones in which we are interested.  (This argument does not directly
use the supersymmetric index of Section \ref{susy-index}.)

Now let us focus on the state content of our protected multiplets, shown in
Figure \ref{two-protected}, and 
compare it with the spectrum of geometric fluctuations of a fivebrane (discussed 
in Section \ref{classical-fivebrane}.)  The expansion of a $2\!\times\! n$-box
superdiagram has been shown in Figure \ref{abcreptwo}. The highest-$\HH$ sector of
the $2\!\times\! n$-box and the lowest-$\HH$ sector of the $2\!\times\! (n-1)$-box diagram
correspond to the fluctuations of a fivebrane along the $x^-$ and radial directions,
whereas the highest weight state of the $2\!\times\! (n-1)$-box diagram and the lowest 
weight state of the $2\!\times\! n$-box diagram correspond to fluctuations along $x^1$ and 
$x^2$.  The remaining fermions match with the fermionic fluctuations of the
fivebrane.  We note that all these states are singlets of $\SO(16)$.

Although we have not identified the complete spectrum of protected
operators on either side, the matching of a natural class of 
protected operators constitutes some evidence that the $(2,2,\dots,2)$ vacuum
describes the transverse fivebrane embedded in the domain wall, similar to
the evidence presented in \cite{tfivebrane} in the case of the BMN matrix model.
We can also make a conjecture about the multi-fivebrane vacua; following
\cite{tfivebrane} it would be natural to consider the vacuum $(2,2,\dots,2;4,4,\dots,4;\dots;2k,2k,\dots,2k)$
as a $k$-fivebrane state.  One could find evidence for this conjecture by looking for $k$ copies of the
spectrum in Figure \ref{two-protected} in this vacuum.

\subsection{The $(1,0)$ superconformal field theory} \label{onezerosection}

In 11-dimensional flat space with a Ho\v{r}ava-Witten domain wall, the low energy
dynamics of a single fivebrane embedded in the domain wall are described by
a $(1,0)$ superconformal field theory in six dimensions.  We have found a vacuum
of our matrix model which is a candidate to describe a large fivebrane embedded 
in the domain wall with topology $S^5$.  We could try to identify the states around this vacuum with
states of the SCFT defined on $S^5 \times \IR$; 
by the state-operator correspondence,
this should give the operator spectrum of that theory, with dimensions $\Delta$ given by the rule 
\begin{equation}
H \leftrightarrow \frac{\mu}{6} \Delta.
\end{equation}
This rule can be justified by the usual logic of the
state-operator correspondence:  namely, when we put the conformal
theory on an $S^5$ of radius $R_5$, we should identify
$R_5 \frac{\partial}{\partial t}$ with $\Delta$, where $t$ denotes
the proper time.  At large $\mu$ the proper time is dominated by
the contribution from $(dx^+)^2$ in the plane wave metric \eqref{planewave},
giving $dt = \frac{\mu}{6} R_5 dx^+$, so that 
\begin{equation}
H = \frac{\partial}{\partial x^+} = \frac{\mu}{6} R_5 \frac{\partial}{\partial t} = \frac{\mu}{6} \Delta,
\end{equation}
as desired.

This identification could also have been made in the case of the BMN matrix model and the 
$(2,0)$ theory.  Encouragingly, it seems to be reasonable there.  Namely, in the $(2,0)$ theory one builds
operators from the scalar fields $X^i$ representing transverse fluctuations, 
of mass dimension $2$, and the derivative operators $\partial^a$ which have
mass dimension $1$; this corresponds to the fact that in the fivebrane vacuum
of the BMN matrix model one has $\beta_{1 m}$ with $H = \frac{\mu}{3}$ and
$x^a_{00}$ with $H = \frac{\mu}{6}$.

So it would be interesting to see whether we can reproduce the operator spectrum of the $(1,0)$
theory using the fivebrane vacuum of our matrix model.  The $(1,0)$ spectrum has been studied
using AdS/CFT in \cite{Gimon} where the operators in short multiplets were classified by their $\SO(4) \times E_8$
quantum numbers.  We can make a few preliminary remarks about this problem.
Indeed, essentially by repeating the procedure from Section
\ref{lambda-spectrum} we can find states which, when completed to
$\SO(4)$ multiplets (this is necessary because the $\SO(4)$ is not manifest
after the pp-wave limit) would be candidates to match operators found in
\cite{Gimon} transforming in the $(\textbf{3}, \textbf{1}, \textbf{248})$ of $\SU(2)_R
\times \SU(2)_L \times E_8$.  These are particularly interesting operators
because they are likely responsible for the transition from the Coulomb
branch to the Higgs branch \cite{GanorHanany,SwittenSix}, where the fivebrane dissolves into a finite
size $E_8$ instanton, producing $30$ hypermultiplets which
parameterize the moduli space of $E_8$ instantons.

However, if indeed our matrix model can describe the $(1,0)$ theory the full story must be subtle, 
because we also find a puzzle:  it seems to be impossible to find the $E_8$
symmetry in the full large $\mu$ spectrum of the matrix model.  In particular, in the 
matrix model we can construct the state $\lambda^r_{-\half} \eta_{\half,-\half} \lambda^s_{-\half} \vac$
which has $H = {7 \mu}/{12}$ and transforms in the $\bf{135 \oplus 1}$ of $\SO(16)$; 
but we have not found a way to complete the $\textbf{135}$ to a multiplet of $E_8$.  The 
smallest candidate is $\textbf{3875}$ but this would require us to find the $\bf{1820 \oplus 1920}$
of $\SO(16)$ elsewhere in the spectrum at $H = {7 \mu}/{12}$, and we have not found a way
to construct these states.  They might arise in a rather complicated way --- already in the membrane
vacuum we had to use both even and odd $N$ sectors to fill out the $E_8$ multiplets, so in the
fivebrane case we might have to combine various vacua which are close to $(2,2,\dots,2)$ in
some appropriate sense.

In sum, further study will be required to see in what sense
there is an $E_8$ symmetry in the fivebrane vacuum of our matrix model and
whether our matrix model can describe the $(1,0)$ theory.

\section{Conclusions and outlook}

In this paper we showed that the plane-wave matrix models may be a
powerful tool to answer many questions in M-theory. A simple system of
harmonic oscillators allowed us to understand, the
$E_8$ symmetry arising from membrane boundaries, the origin
of giant gluons and anomaly cancellation in
the membrane worldvolume; furthermore it seems to capture at least
some of the degrees of freedom of the fivebrane embedded in the 
domain wall, and we may hope that indeed it will capture all of them.
Clearly, we have not resolved all outstanding questions about heterotic M-theory.
In particular, we have failed to clarify
the following issues in the plane-wave matrix models, which in our opinion
deserve further investigation:

\begin{itemize}
\item {\bf Understanding the $\mu\to 0$ limit.}
The $\mu\to 0$ limit is a highly nonperturbative regime from the viewpoint 
of the perturbative expansion of the plane-wave matrix model. Some states
that are guaranteed to exist for finite $\mu$ become non-normalizable at 
$\mu=0$. It is desirable to find some machinery that allows us to prove 
that the ``right'' states survive and the others don't. A full argument 
could use the supersymmetric indices combined with the $\SO(8)$ rotational 
symmetry that gets restored for $\mu=0$ (or $\SO(9)$ in the case of the 
BMN model).

\item {\bf The spectrum of operators in the $(1,0)$ theory.} 
As we discussed in Section \ref{onezerosection}, using
the state-operator correspondence
it might be possible to identify explicitly the operator spectrum 
of the $(1,0)$ theory by a careful study of the spectrum of
our matrix model around the fivebrane vacuum.
Because the plane-wave background breaks some of the symmetries, not all 
components of the R-symmetry multiplets survive to $\mu=\infty$. In our 
case, the $\SO(4)$ R-symmetry was broken to $\U(1)$ generated by $M^{12}$. 
Similarly, the $\SO(5)$ R-symmetry of the $(2,0)$ theory was broken to 
$\SO(3)$ in the BMN case.  It might be possible to find a way to identify 
the rest of the operator spectrum or at least a well-defined rule that would 
determine which operators are visible in the matrix model and how many 
``invisible'' (i.e.\ unprotected) operators there are.  One might also try
to understand whether there is some relation between our matrix model, in which the
fivebrane arises dynamically, and the matrix models describing longitudinal
flat fivebranes \cite{twozeromatrix,onezeromatrix}.  Finally, it might be possible
to exhibit explicitly the $E_8$ symmetry which should be there in the $(1,0)$ theory,
although as discussed in Section \ref{onezerosection} this will apparently require some new idea.

\item {\bf A more complete proof of the $E_8$ symmetry for membranes
at large $N$.} We were able to 
show that the membrane spectrum at $\mu=\infty$ forms full representations 
of $E_8$.  A more complete proof for finite $\mu$, including the 
interactions of membranes, might still result from an application of the 
techniques of two-dimensional CFT's, in agreement with the dimension of 
the boundaries.

\item {\bf More general backgrounds.} There might be other backgrounds of 
M-theory that admit a matrix model description. For example, it might be 
interesting to study various supersymmetric orbifolds of the BMN plane 
wave, i.e.\ the massive deformations of the ALE spaces, as well as the 
matrix models for stringy pp-wave backgrounds.

\end{itemize}

Although the ultimate reach of these matrix models will most likely be limited to 
highly symmetric backgrounds similar to ours, we are hopeful that some general insights 
resulting from these models might have a broader range of validity.

\acknowledgments

We would like to express our special gratitude to 
Michal Fabinger for his collaboration at the early
stages of this work.  We are also grateful to
Nima Arkani-Hamed,
Keshav Dasgupta, 
Eric Gimon, 
Shiraz Minwalla,
Mark Van Raamsdonk, 
and
Andrew Strominger
for very useful discussions. This work was supported in part by Caltech
DOE grant DE-FG03-92-ER40701, Harvard DOE grant DE-FG01-91ER40654
and the Harvard Society of Fellows. The work of M. M. Sh-J. is   
supported in part by NSF grant PHY-9870115 and in part by funds from the
Stanford Institute for Theoretical Physics.  The work of A. N. is supported by
an NDSEG Graduate Fellowship.

\appendix

\section{Two-oscillator states around the fivebrane vacuum} \label{two-osc-appendix}

The one-oscillator states around the $(2,2,\dots,2)$ vacuum were discussed
in the main text, in Section \ref{fivebrane-spectrum}.
We may similarly analyze the two-oscillator states, obtained by
symmetric tensor multiplication of two single-oscillator superdiagrams.
As an example we show the symmetric product of two such diagrams:

\vspace{3mm}

\epsfig{file=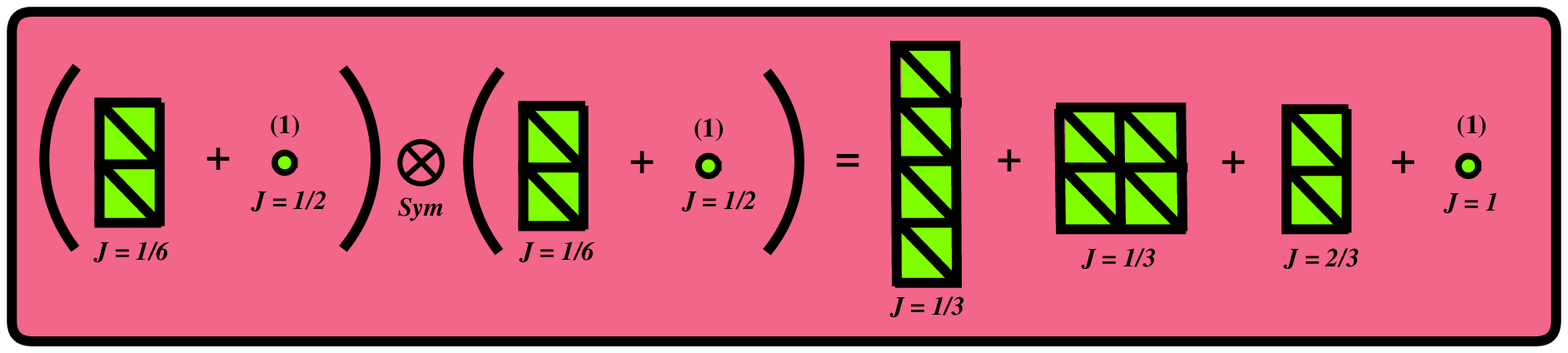,width=140mm}

where

\vspace{3mm}

\epsfig{file=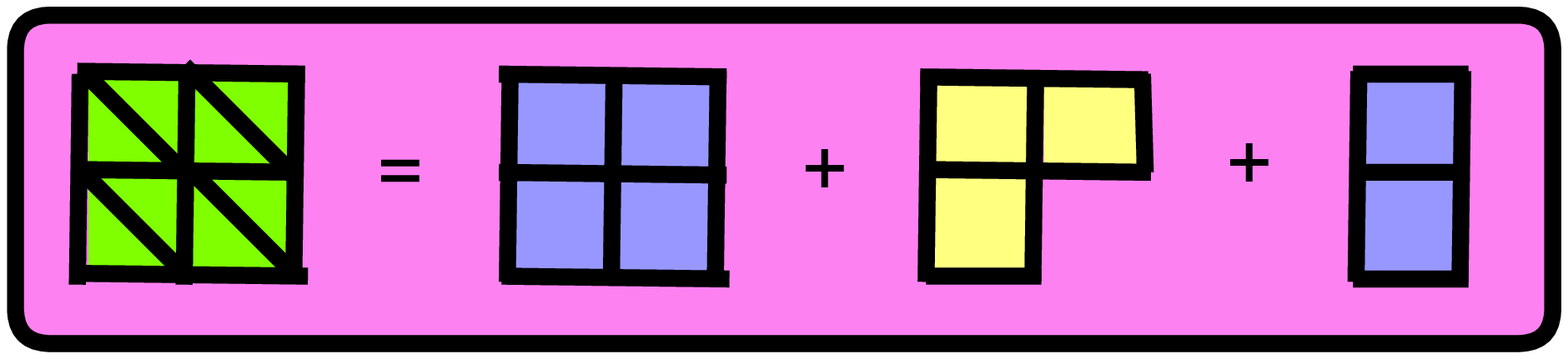,width=85mm}

The first multiplet in the two-oscillator state 
decomposition is typical, while 
the other three are atypical.\footnote{In general, any 
tensor product of irreducibles involving a typical superdiagram  
only decomposes into typical representations.  One can roughly understand this by noticing
that whenever one of the representations we start with is
typical, i.e.\ $B=F$, then the tensor product will also have $B=F$.} 

Equipped with superdiagram multiplications, one can easily work out the 
two-oscillator states. The two oscillators may be chosen from Tables 
\ref{osc2-1} or \ref{osc2-2}.  All the two-oscillator 
states in which both oscillators come from Table 
\ref{osc2-1} are shown in Figure \ref{figureTWO-OSCILU(1)}.

\EPSFIGURE{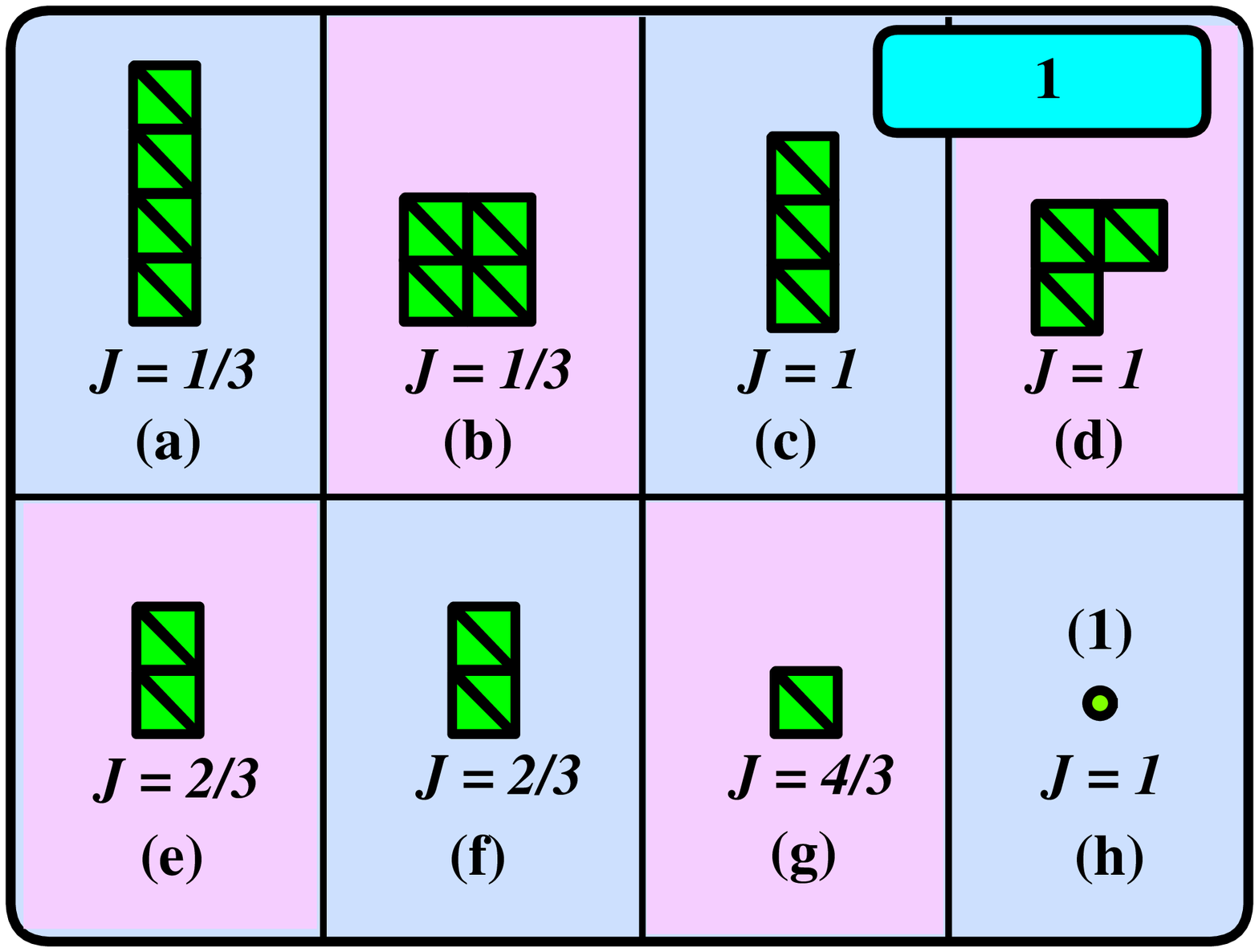,width=105mm}{Two
oscillator states of the $\U(1) \subset \U(2)$ sector of the $\OO(2N)$
theory which are $\SO(16)$ singlets. The highest weight 
state of these multiplets are:
{\bf (a)} $\Tr (x^a_{00}x^a_{00})|0\rangle$; 
{\bf (b)} $\Tr 
(x^a_{00}x^b_{00}-\frac{1}{6}\delta_{ab}x^c_{00}x^c_{00})|0\rangle$; 
{\bf (c,d)} $\lambda^r \eta_{\frac{1}{2},-\frac{1}{2}}x^a_{00}\lambda^r 
|0\rangle$; 
{\bf (e)} $\Tr 
(\eta_{\frac{1}{2},-\frac{1}{2}}\eta_{\frac{1}{2},-\frac{1}{2}})|0\rangle$; 
{\bf (f)} $\Tr (x^a_{00}\beta_{1,-1})|0\rangle$; 
{\bf (g)} $\lambda^r \eta_{\frac{1}{2},-\frac{1}{2}}\beta_{1,-1}\lambda^r 
|0\rangle$;  
{\bf (h)} $\Tr (\beta_{1,-1}\beta_{1,-1})|0\rangle$. 
Note that except {\bf (a)} all the 
multiplets are atypical. \label{figureTWO-OSCILU(1)}}

\EPSFIGURE{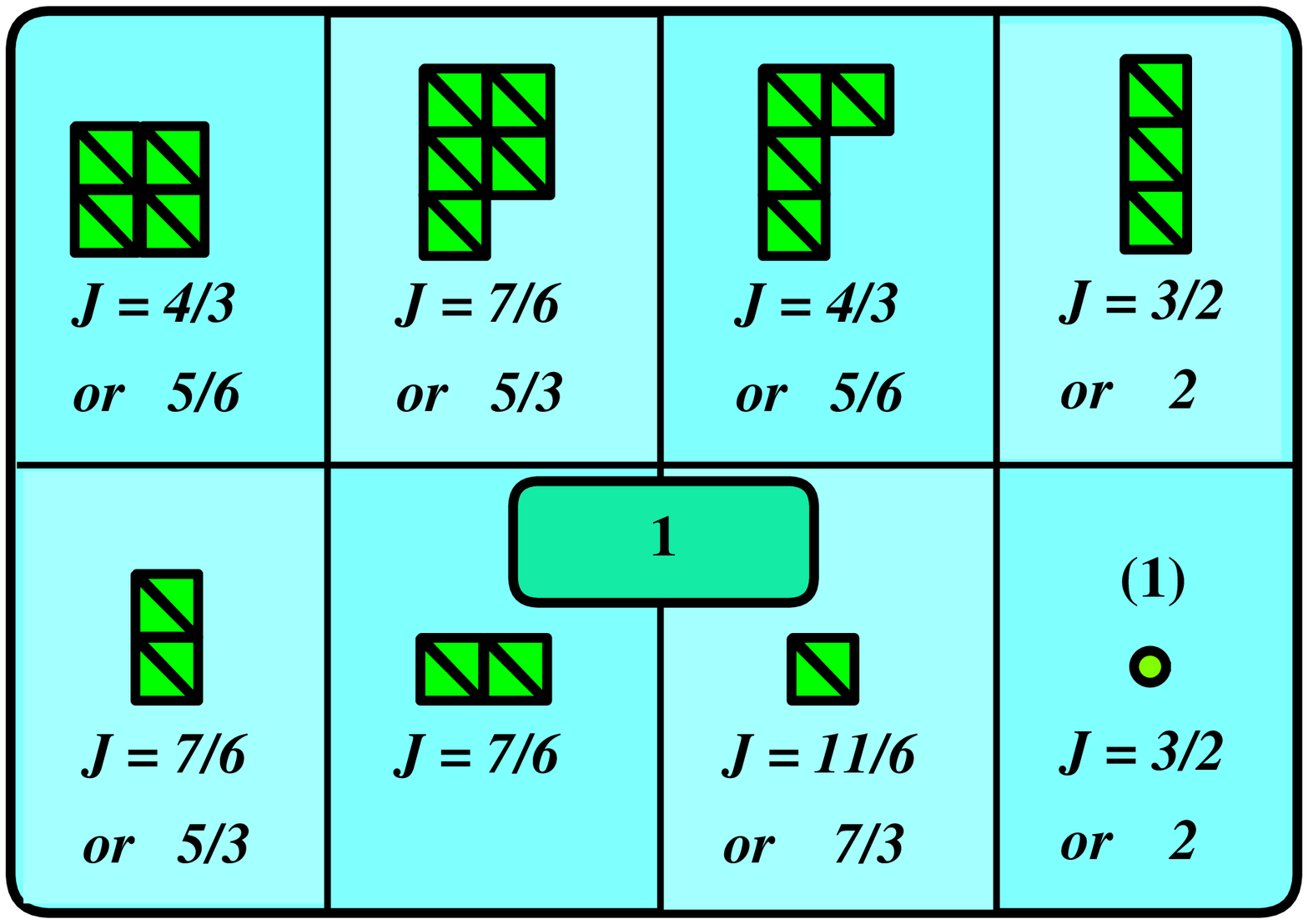,width=115mm}{Two
oscillator atypical multiplets of the $\OO(2N)$
theory which are $\SO(16)$ singlets. \label{figureTWOOSCIL}}

One may similarly construct states involving one oscillator of Table 
\ref{osc2-1} and one oscillator of Table \ref{osc2-2} or 
two oscillators of Table \ref{osc2-2}. These states may be in ${\bf 
1, 120}$ or ${\bf 135}$ of $\SO(16)$. In Figure \ref{figureTWOOSCIL}
we have gathered the {\it atypical} multiplets from the rest of the two-oscillator states 
about the $(2,2,\cdots, 2)$ vacuum which are $\SO(16)$ singlets (several
more two-oscillator atypicals have already been depicted in Figure \ref{figureTWO-OSCILU(1)}).

As we see from Figures \ref{figureTWO-OSCILU(1)} and \ref{figureTWOOSCIL}, 
many of these atypical representations
can find partners to combine with and receive energy 
shifts. The $2\times 2$-box diagram with $J={4}/{3}$ can also find its partner 
(which is the $(0,2,1|0)$ multiplet) among the three oscillator states and hence 
it is also not protected.

The above discussion generalizes in an obvious way to $m$-oscillator
states; we find typical multiplets as well as atypical ones; for example, some of 
the $n+k$-oscillator states which are relevant for the fivebrane
interpretation have been shown in Figure \ref{fig-1}.

\EPSFIGURE{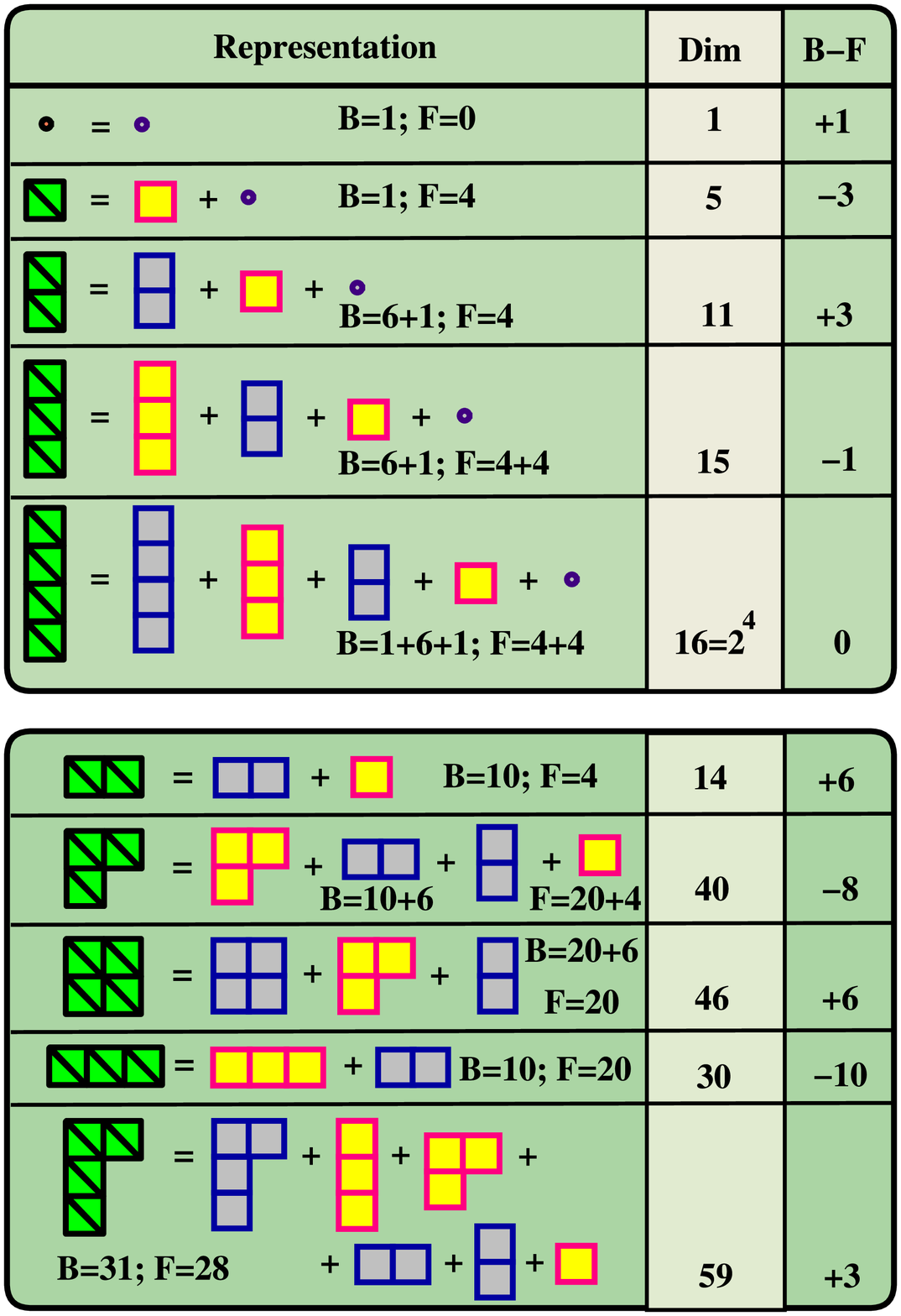,width=130mm}{Some examples
of the simplest $\SU(4|1)$ superdiagrams and their
decomposition under $\SU(4)$. The slashed boxes
represent superindices while the ordinary boxes are $\SU(4)$
indices. Bosonic representations are shaded in grey and fermionic
representations are shaded yellow.
\label{abcrepone}}

\EPSFIGURE{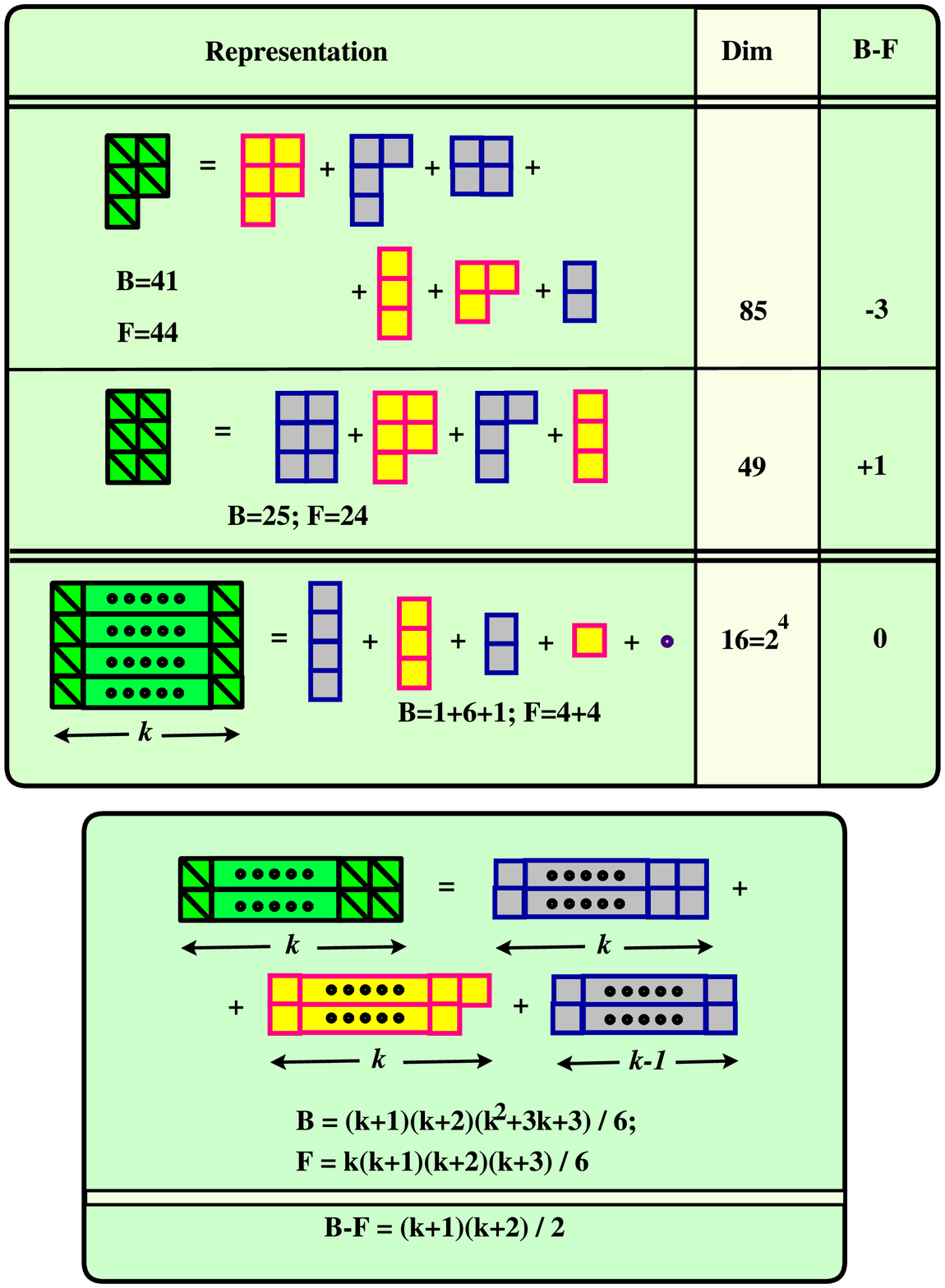,width=145mm}{Some more examples
of the simplest $\SU(4|1)$ superdiagrams.
\label{abcreptwo}}

\end{document}